\documentclass[conference]{IEEEtran}
\usepackage{amsmath, amssymb}
\usepackage{graphicx}
\usepackage{float}
\usepackage{caption}
\usepackage{subcaption}
\usepackage{hyperref}
\usepackage{geometry}
\usepackage{booktabs}
\usepackage{tabularx}
\usepackage{lscape}
\usepackage{enumitem}
\usepackage{tikz}
\usepackage{titlesec}
\usepackage{verbatim}
\usepackage{multirow}
\usepackage{siunitx}
\usepackage{makecell}
\usepackage[flushleft]{threeparttable}
\usepackage[font=small]{caption}

\usepackage{fancyhdr}

\usepackage{hyperref}  

\title{AI Load Dynamics--A Power Electronics Perspective}

\author{%
Yuzhuo Li\textsuperscript{1},
and~Yunwei Li\textsuperscript{1}
\\\textsuperscript{1}Department of Electrical and Computer Engineering\\
University of Alberta, Edmonton, Canada\\
Email: \texttt{yuzhuo@ualberta.ca}, \texttt{yunwei.li@ualberta.ca}
}

\date{Jan. 2025}

\begin{document}
\maketitle

\begin{abstract}
As AI-driven computing infrastructures rapidly scale, discussions around data center design often emphasize energy consumption, water and electricity usage, workload scheduling, and thermal management. However, these perspectives often overlook the critical interplay between AI-specific load transients and power electronics. This paper addresses that gap by examining how large-scale AI workloads impose unique demands on power conversion chains and, in turn, how the power electronics themselves shape the dynamic behavior of AI-based infrastructure. We illustrate the fundamental constraints imposed by multi-stage power conversion architectures and highlight the key role of final-stage modules in defining realistic power slew rates for GPU clusters. Our analysis shows that traditional designs, optimized for slower-varying or CPU-centric workloads, may not adequately accommodate the rapid load ramps and drops characteristic of AI accelerators. To bridge this gap, we present insights into advanced converter topologies, hierarchical control methods, and energy buffering techniques that collectively enable robust and efficient power delivery. By emphasizing the bidirectional influence between AI workloads and power electronics, we hope this work can set a good starting point and offer practical design considerations to ensure future exascale-capable data centers can meet the stringent performance, reliability, and scalability requirements of next-generation AI deployments.\end{abstract}

\section{Introduction}

The rapid evolution of artificial intelligence (AI) workloads has fundamentally transformed the landscape of data center power infrastructure \cite{doe2024poweringAI,abb2024datacenter}. 
Contemporary AI training clusters (for Large Language Models (LLMs) from OpenAI, Google, Meta, etc.) now routinely consume power at megawatt scales (or even beyond) while exhibiting unprecedented power dynamics during operation \cite{li2024unseenaidisruptionspower}. This paradigm shift necessitates a comprehensive re-examination of power delivery architectures, with particular emphasis on the fundamental limitations imposed by cascaded power conversion chains. Understanding these constraints is critical not only for current deployments but also for the strategic development of next-generation AI infrastructure.

\textbf{Research Questions:}
\begin{itemize}
    \item What is the fundamental timescale bottleneck in AI data center power chains?
    \item How does each power conversion stage contribute to limiting overall power ramp rates in Graphics Processing Unit (GPU) clusters?
\end{itemize}

\begin{figure}[h]
  \centering
  \includegraphics[width=0.7\columnwidth]{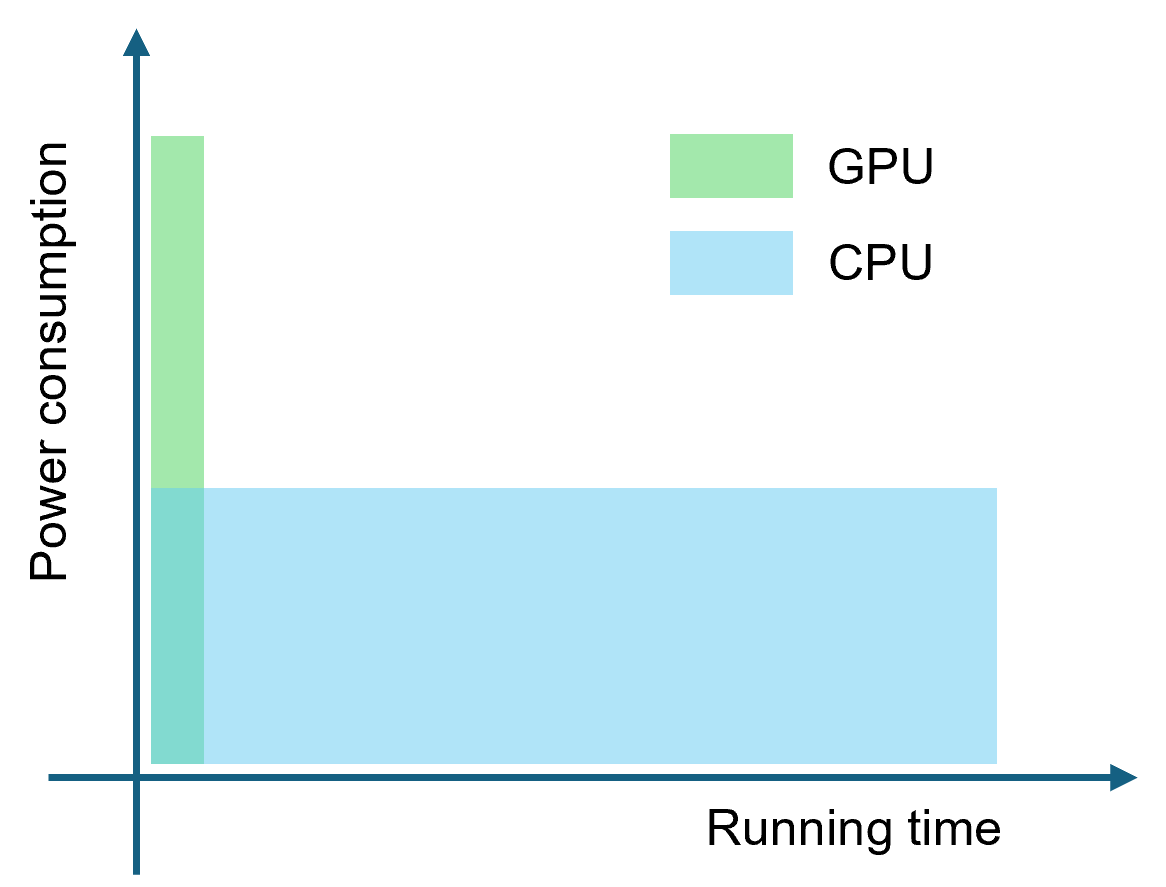}
  \caption{Different Power Consumption Features of GPU vs. CPU Operation for same AI Computation Task.}
  \label{fig:GPU_CPU}
\end{figure}

The scale and sophistication of AI workloads continues to accelerate, straining both electricity supply and facility design. Projections indicate that AI could drive a steep surge in U.S. data-center power demand in the coming decade \cite{goldman2024AIpower}.
Indeed, EPRI’s comprehensive analysis shows that AI-centric infrastructures can significantly elevate overall power consumption \cite{epri2024poweringintelligence}. Training clusters have evolved from traditional configurations consuming 10-20kW per rack to advanced designs exceeding 100kW and even reach to 350kW per cabinet. This progression reflects both the increasing computational demands of modern AI algorithms and the relentless pursuit of improved training efficiency. The economic implications are substantial, with power infrastructure representing a significant portion of total data center capital expenditure. Moreover, the dynamic nature of AI workloads—characterized by rapid transitions between computational phases—places unprecedented demands on power delivery systems. 
As shown in Fig.~\ref{fig:GPU_CPU}, GPUs have a different power profile compared to traditional Central Processing Units (CPUs), further emphasizing the unique power dynamics AI accelerators introduce. Empirical measurements also confirm that heterogeneous AI workloads can easily drive dynamic power swings  \cite{caspart2022preciseenergyconsumptionmeasurements}.

Current industry practice reveals several critical challenges. First, the complex interaction between multiple power conversion stages, spanning facility-level AC-DC conversion to point-of-load regulation, creates cascaded dynamic limitations that constrain system response capabilities. Second, the need for sophisticated energy storage distribution and protection coordination schemes introduces additional complexity and potential bottlenecks. Third, the integration of advanced cooling solutions requires careful consideration of power delivery system dynamics.

These challenges are exacerbated by the rapid pace of AI hardware evolution. The transition from traditional CPU or CPU/GPU architectures to specialized AI accelerators has driven substantial increases in both power density and dynamic range. For instance, modern AI accelerators can exhibit power variations exceeding 50\% of their thermal design power (TDP) within milliseconds, while next-generation devices are expected to push these boundaries even further. This trend necessitates sophisticated power delivery architectures capable of handling both steady-state power requirements and rapid transient responses.

Understanding and optimizing these power dynamics requires careful consideration of multiple interacting constraints. Power conversion chains and workload patterns contribute significantly to system response capabilities. This interaction of multiple constraints has profound implications for the design and operation of AI infrastructure, particularly as the industry moves toward ever-larger training clusters and more dynamic workload patterns.

\section{GPU Rack/Cluster Hardware}

The evolution of GPU cluster configurations demonstrates a consistent trajectory toward increased power density capabilities, presenting significant challenges in data center design and operation. Contemporary deployments span a broad spectrum from traditional several-kW compute racks to advanced AI-ready solutions exceeding 100kW power consumption per rack \cite{balaban2020gpucluster}. Industrial reference guides further detail advanced cooling solutions and validated power-distribution topologies \cite{drivenets_ai_cluster_guide,schneider_nvidia_ai_reference_designs}. This section presents a systematic analysis of these configurations, examining their architectural characteristics and operational constraints across three primary scales of implementation.

\subsection{Single-Rack Architectures}

\begin{table*}[t]
\centering
\caption{GPU Cluster Power Configurations}
\label{tab:confirmed_power_configs}
\begin{tabular}{l|c|c|c|c}
\toprule
\textbf{Configuration} & \textbf{Total Power} & \textbf{GPU Type} & \textbf{GPU Power} & \textbf{System Density} \\
\midrule
\multicolumn{5}{l}{\textbf{Single-Rack Configurations}} \\
\midrule
Traditional A100 & 32\,kW & NVIDIA A100 & 400\,W/GPU & 8 servers × 8 GPUs = 64 GPUs \\
\midrule
Traditional H100 & 34\,kW & NVIDIA H100 & 700\,W/GPU & 6 servers × 8 GPUs = 48 GPUs \\
\midrule
NVL36 Liquid-Cooled & 73\,kW & GB200 & 1,200\,W/GPU & 9 servers × 4 GPUs = 36 GPUs \\
\midrule
NVL72 Liquid-Cooled & 132\,kW & GB200 & 1,200\,W/GPU & 18 servers × 4 GPUs = 72 GPUs \\
\bottomrule
\end{tabular}
\end{table*}

\subsubsection{NVL Series Implementation}
The NVL36 configuration, with up to 2700W for the complete Grace Blackwell Superchip (GB200), implements a comprehensive design supporting 9 servers with 4 GPUs each. This architecture incorporates enhanced power delivery systems integrated with Cooling Distribution Units (CDUs), achieving a total rack power capacity of 73kW. Building upon this foundation, the NVL72 configuration doubles the compute density to accommodate 18 servers with 4 GPUs each. This advanced design implements dual power delivery paths, enhancing system reliability through sophisticated power management systems capable of handling 132kW per rack.

\subsubsection{Ultra-High-Density Implementations}
The ultra-high-density configuration, exemplified by the HPE Cray EX platform \cite{hpe_cray_supercomputing_ex}, represents another current state-of-the-art in GPU computing infrastructure. This implementation supports an unprecedented density of 224 GPUs per cabinet with a total power capacity of 350kW. The system architecture comprises eight compute chassis per cabinet, each supporting eight compute blade slots, complemented by four redundant power shelves with dedicated PDUs in a quad power input configuration.

The node implementation in these systems utilizes the EX254n blade architecture, featuring a dual-node design that integrates four GH200 superchips per node. Each node incorporates a 72-core Arm Neoverse V2 Grace CPU, optimized with 128GB LPDDR5X DRAM, and quad Slingshot-11 Network Interface Card (NIC) implementation. This configuration achieves maximum density while maintaining thermal stability through advanced management systems.

\subsection{Multi-Rack Industrial Reference Designs}

\subsubsection{Reference Architecture Overview}
Industrial-scale GPU reference designs, such as Schneider’s EcoStruxure RD109 (7.392\,MW for IT load) \cite{ecostruxure_design_109}, demonstrate how hierarchical power distribution and advanced cooling can be integrated. 
Similarly, Vertiv’s 360AI concept outlines multi-rack scaling for AI infrastructures \cite{vertiv_360ai_brochure}. These reference designs provide validated architectures for scalable AI infrastructure deployment, incorporating comprehensive solutions for power delivery and system management.

\begin{figure*}[t]
  \centering
  \includegraphics[width=0.7\textwidth]{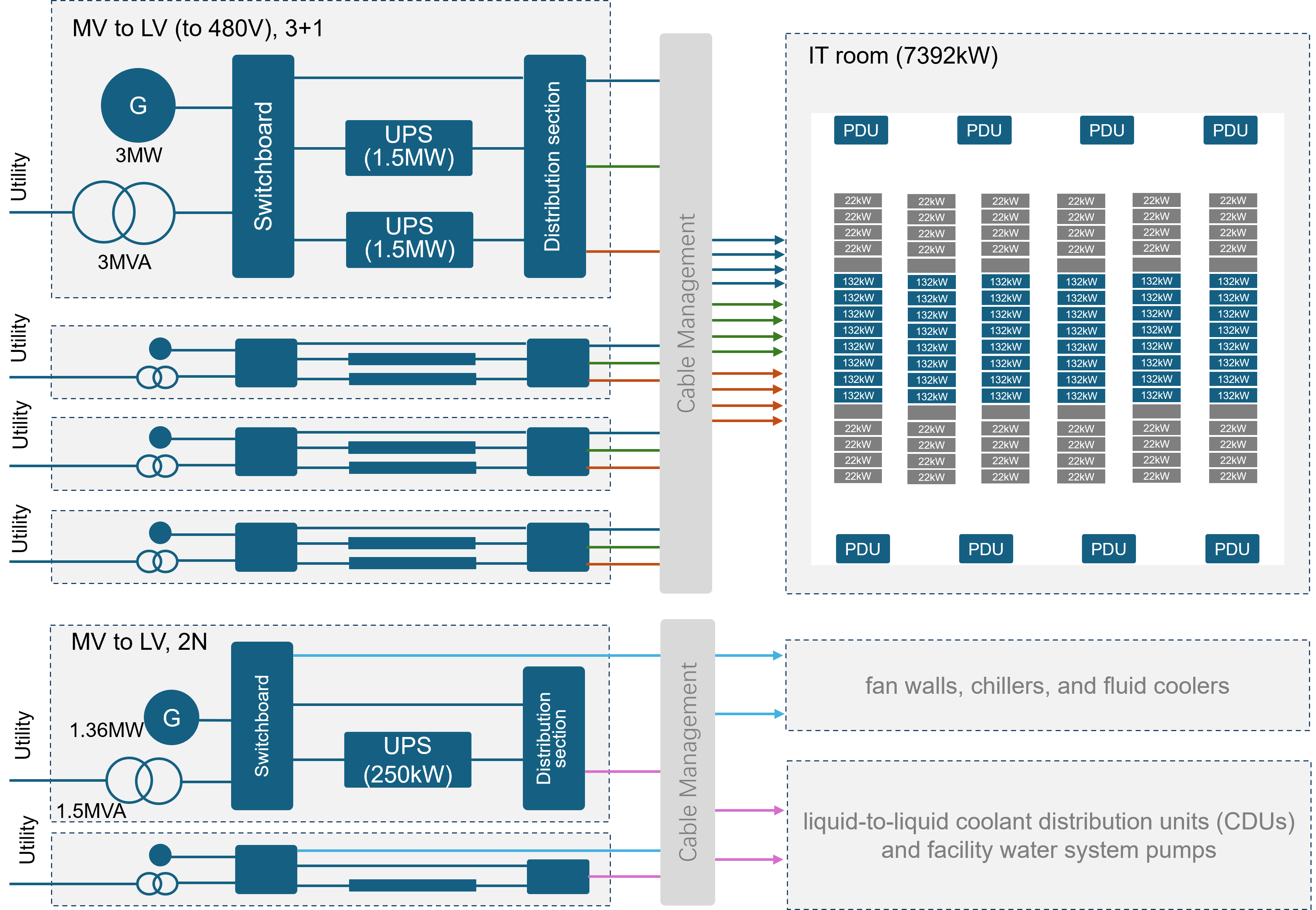}
  \caption{GPU cluster reference design implementation demonstrating comprehensive power distribution architecture (from Schneider RD109, 7392kW for IT racks, total power consumption approximately 10.5MW including cooling infrastructure)}
  \label{fig:Power_chain_ref}
\end{figure*}

\subsubsection{Power Distribution Architecture}
\begin{figure}[t]
  \centering
  \includegraphics[width=\columnwidth]{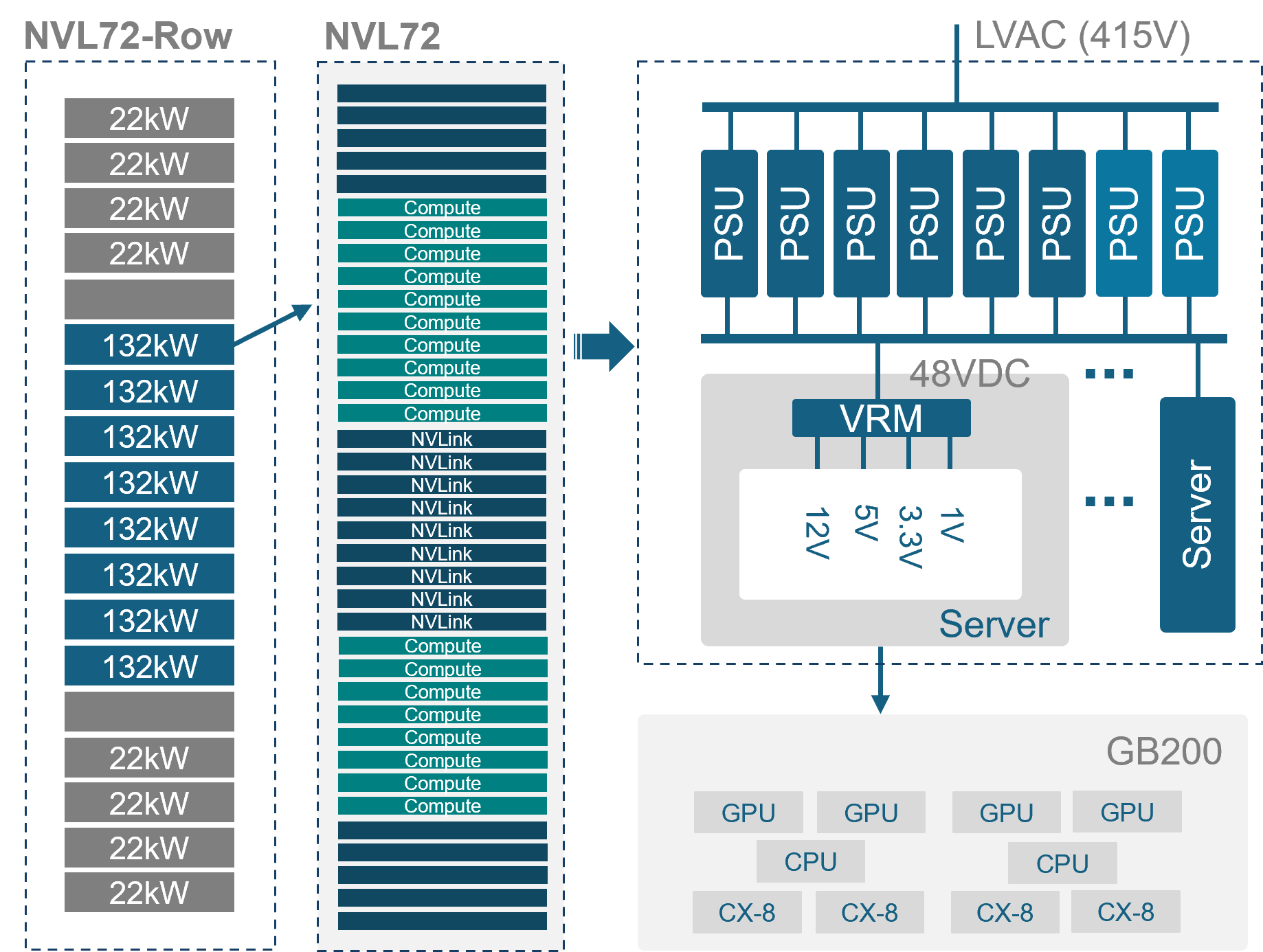}
  \caption{A possible NVL72 rack architecture showcasing integrated compute and power distribution systems}
  \label{fig:NVL72_rack}
\end{figure}

Figure~\ref{fig:Power_chain_ref} shows a representative industrial reference design featuring hierarchical power distribution and cooling strategies. The power architecture in multi-rack implementations demonstrates sophisticated hierarchical organization. The primary power distribution infrastructure utilizes Medium-Voltage AC (MVAC) utility feeds, complemented by parallel 3MVA transformers and 3MW generator systems per power train. The system incorporates N+1 redundant medium voltage switchgear and a 480V Low-Voltage AC (LVAC) distribution backbone organized in four parallel 3MW power trains configured in a 3+1 distributed redundant arrangement. And the major computation racks feature NVL72 architecture as detailed in Figure~\ref{fig:NVL72_rack}.

\subsubsection{Implementation Scales}
The 904kW configuration exemplifies sophisticated system integration methodologies, combining eight compute racks at 73kW each with eight network infrastructure racks at 40kW each. Scaling this architecture, the 1808kW configuration demonstrates comprehensive multi-rack implementation strategies, incorporating sixteen compute racks at 73kW each alongside sixteen network infrastructure racks. The 7392kW implementation represents a state-of-the-art high-density GPU cluster architecture utilizing sophisticated multi-tier power distribution topology. 

\subsection{Large-Scale Deployment Projections}

\subsubsection{Workload-Specific Architectures}
Deployments spanning from 10,000 up to 1,000,000 GPUs require rethinking both facility power and data pipelines. Additionally, some workloads are shifting to smaller edge sites \cite{vertiv2019closertotheedge}, while enterprise High-Performance Computing (HPC) orchestration must unify large AI workflows 
\cite{hammerspace_ai_workflows}.
Training-intensive deployments demonstrate distinctive requirements in power infrastructure design, implementing sustained high-power operation optimization with direct liquid cooling solutions predominating the thermal management strategy. These systems typically employ 3+1 or 4-to-make-3 redundancy schemas complemented by sophisticated management systems. 

Inference-focused deployments present contrasting architectural requirements, emphasizing dynamic load profile management capabilities through hybrid cooling implementations. These systems typically employ N+1 redundancy schemas with rapid failover mechanisms, supported by advanced load balancing systems. System characteristics reflect variable workload patterns, requiring sophisticated variable thermal load management systems and network architectures optimized for low-latency response. Storage subsystems are engineered specifically for high Input/Output operations per second performance to support rapid data access patterns characteristic of inference operations.

Hybrid architectures must accommodate training-intensive racks where sustained high-power operation dominates, as well as dynamic inference loads prone to frequent ramp-ups \cite{gu2023energyefficientgpuclustersscheduling}. 
It requires complex power and cooling infrastructures capable of accommodating diverse workload characteristics. These systems implement multi-tier power distribution topologies featuring modular power subsystems and dynamic power allocation mechanisms. The thermal management integration combines multiple cooling approaches, incorporating variable thermal load handling capabilities with Cooling Distribution Unit capacity optimization algorithms. Temperature gradient management systems ensure uniform thermal conditions across the deployment, critical for maintaining consistent performance characteristics under varying workload conditions.

\section{Workload Pattern Characterization}
\label{sec:ai_workloads}



Modern AI workloads present unique challenges for data center power infrastructure that fundamentally differ from traditional computing loads. To understand these challenges, we must first examine how AI accelerators like GPUs operate and why their power consumption patterns are distinctive. This section provides a comprehensive analysis of these patterns and their implications for power system design. All experimental waveforms presented in this section were captured on a single in-lab workstation running Ubuntu~20.04 LTS with CUDA~12.x, equipped with an AMD~Ryzen~5~5500 (3.6\,GHz base) processor, 32\,GB DDR4 (2\,×\,16\,GB G.Skill RipjawsV at 3200\,MT/s), an NVIDIA~RTX\,4090 GPU (standard reference design with a 16-pin PCIe~5.0 connector), and a 1000\,W MPG PCIe~5.0 Gold (80+ Gold) power supply unit. The workload software included two primary AI models—\textbf{GPT-2~(124\,M parameters, a Generative Pretrained Transformer model from OpenAI)} under PyTorch for training experiments and \textbf{LLaMA-3.1~(8\,B parameters, LLM model from Meta)} for inference tests via a custom script. A Tektronix~DPO-series oscilloscope, configured with Hall-effect current probes on the GPU motherboard (Ch1) and Power Supply Unit (PSU) input (Ch2) alongside single-phase~AC voltage monitoring (Ch3), recorded all waveforms. Each figure in this section provides time-synchronized captures over various zoom levels, enabling detailed observation of the GPU’s rapid load transitions.

\subsection{Transient Measurements}

\begin{figure*}[t]
    \centering
    \begin{subfigure}[b]{0.48\textwidth}
        \centering
        \includegraphics[width=\textwidth]{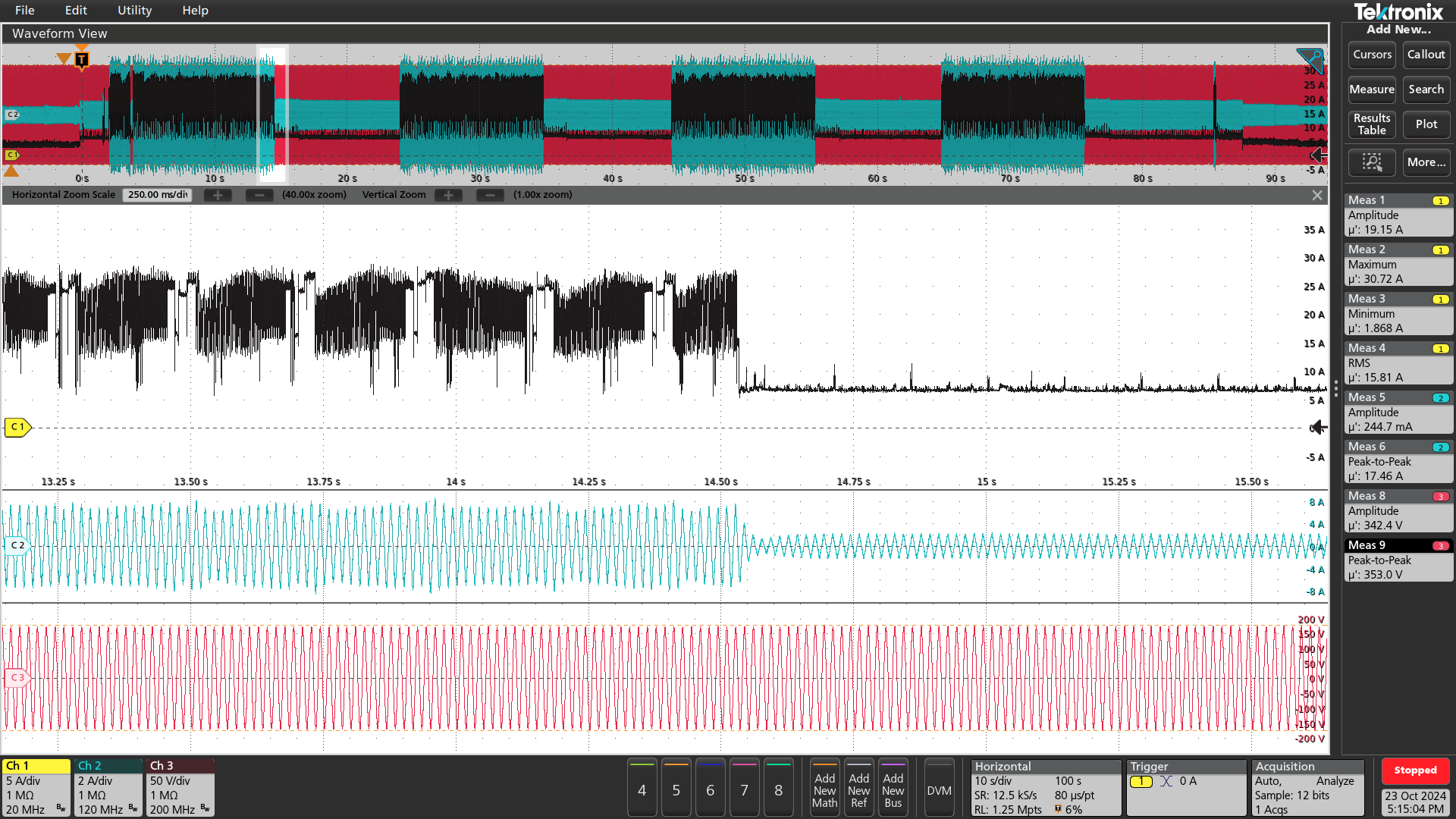}
        \caption{Initial capture showing the first load drop around 14.5\,s. 
        Checkpoints induce a sudden drop from peak current to near-idle, 
        underscoring the abrupt nature of AI training workload transitions.}
        \label{fig:first_load_drop}
    \end{subfigure}
    \hfill
    \begin{subfigure}[b]{0.48\textwidth}
        \centering
        \includegraphics[width=\textwidth]{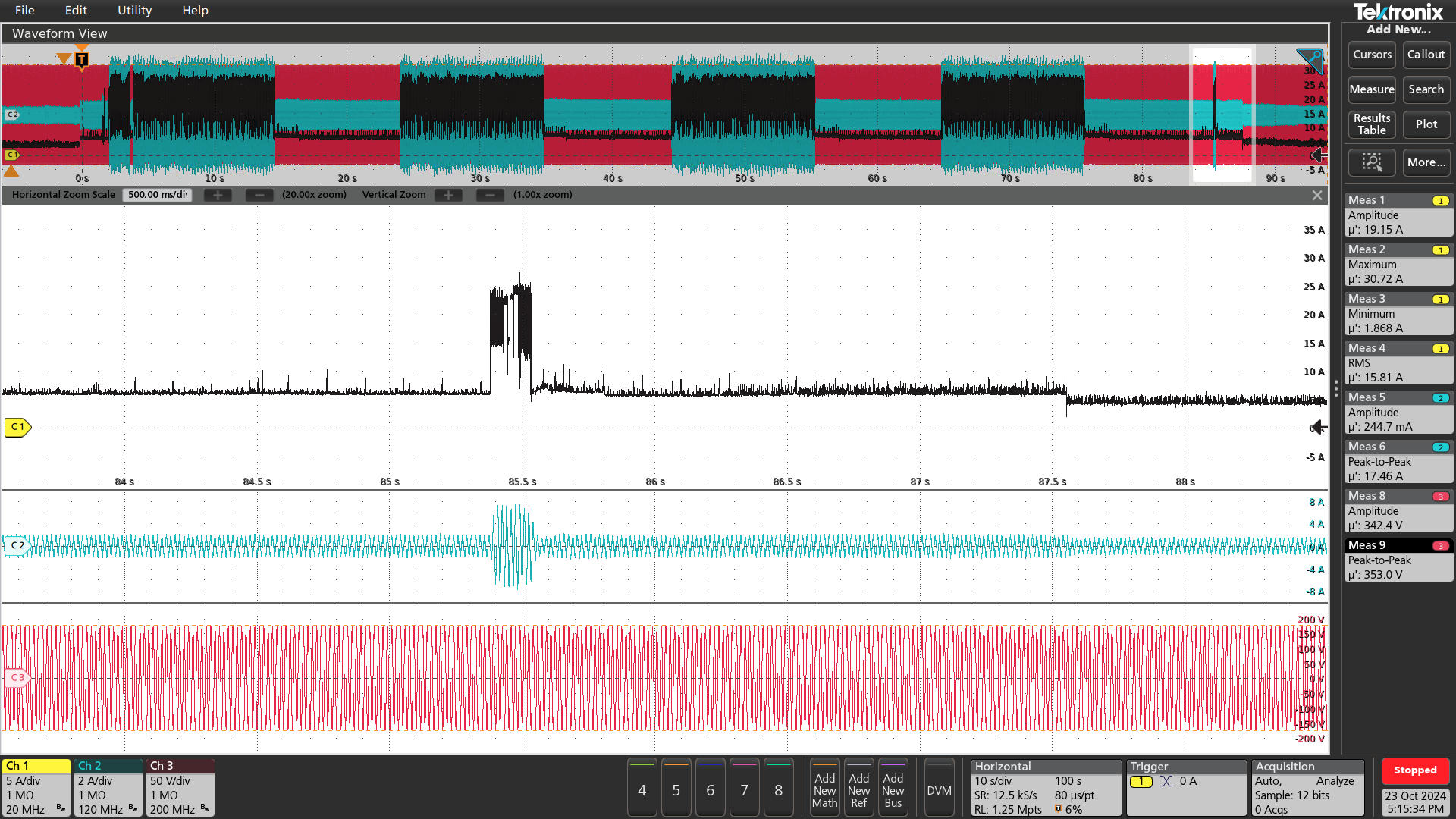}
        \caption{Closer view (10\,s scale) of the last checkpoint event, 
        where training is intentionally interrupted, causing rapid up/down current surges.}
        \label{fig:final_zoom_10s}
    \end{subfigure}

    \vspace{1em} 

    \begin{subfigure}[b]{0.48\textwidth}
        \centering
        \includegraphics[width=\textwidth]{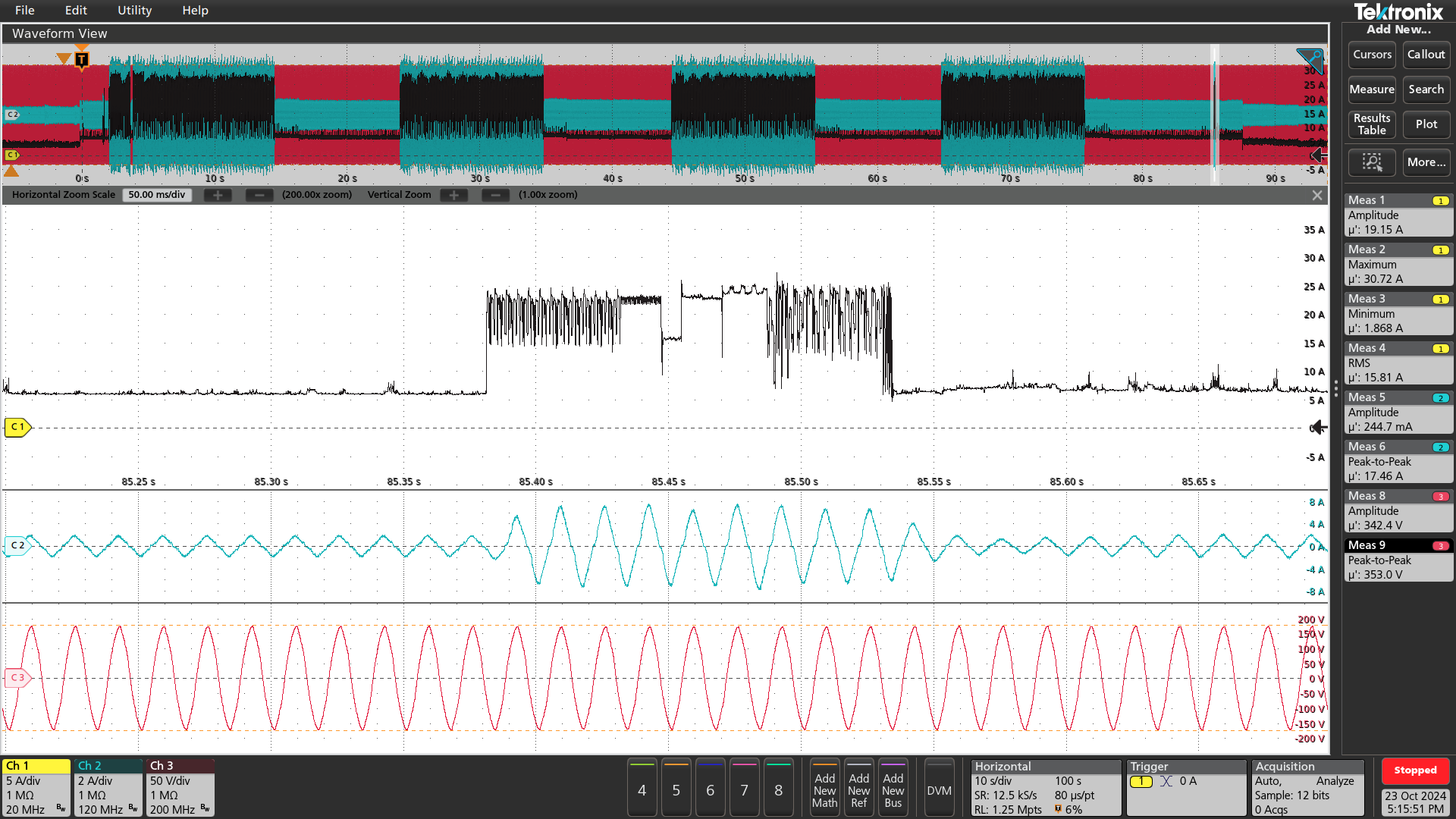}
        \caption{Detailed capture (500\,ms scale) focusing on the steep transitions 
        as the GPU goes from full compute load to near idle. Several fundamental AC cycles 
        are visible in the PSU current waveform.}
        \label{fig:final_zoom_500ms}
    \end{subfigure}
    \hfill
    \begin{subfigure}[b]{0.48\textwidth}
        \centering
        \includegraphics[width=\textwidth]{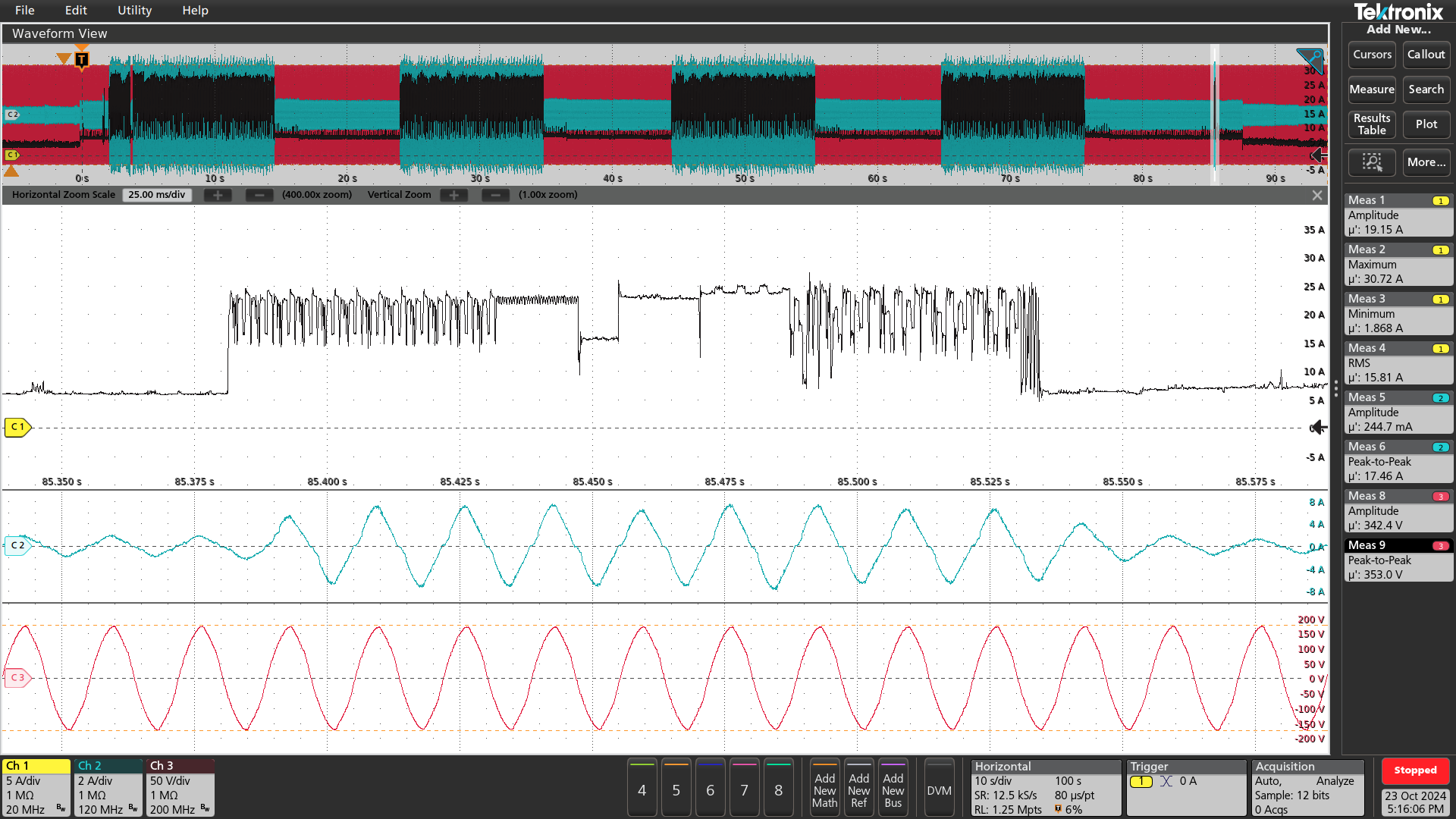}
        \caption{Millisecond-level zoom of the final interrupt event, pinpointing 
        the instantaneous swings in GPU motherboard current 
        across just a few AC cycles.}
        \label{fig:final_zoom_ms}
    \end{subfigure}

    \caption{Progressive zoom-in of GPU current transients during GPT-2 (124\,M) training checkpoints.
    (a) illustrates the first load drop, while (b)--(d) progressively zoom into the last 
    interrupted training phase, where rapid up/down surges occur 
    within a span of several AC cycles. Waveform from top to bottom: current of the GPU mother board, PSU current, PSU voltage.}
    \label{fig:gpu_transient_zoom}
\end{figure*}

\paragraph{During GPT-2 Training Checkpoints}
As illustrated in Figure~\ref{fig:gpu_transient_zoom}, the GPU motherboard current experiences abrupt, multi-ampere surges associated with model checkpoint events. Subfigure~\ref{fig:first_load_drop} captures an early load drop around 14.5\,s; one can observe the current plunging from near-peak values to an almost idle state within milliseconds. This sudden negative transient underscores the need for fast PSU control to avert overvoltage on the internal DC bus. Meanwhile, Subfigure~\ref{fig:final_zoom_10s} zooms to the final checkpoint event at a 10\,s timescale, showing rapid current swings when the training process is intentionally halted. By narrowing the view further to 500\,ms in Subfigure~\ref{fig:final_zoom_500ms} and eventually to a millisecond-level capture in Subfigure~\ref{fig:final_zoom_ms}, we see how swiftly the GPU transitions from full compute load to near idle across merely a few AC cycles.

Although local energy buffering partially smooths these spikes, the magnitude and speed of these current changes go well beyond traditional CPU-centric or transaction-based workloads. When extrapolated to multi-rack or multi-megawatt HPC clusters, such fast load fluctuations can stress upstream power distribution, control loops, and protective devices. Hence, GPU-centric systems demand not only higher average power but also the capability to handle large di/dt events on millisecond timescales, as predicted by the formal analysis in Section~\ref{sec:power_chain_dynamics}.

\noindent
\textbf{Comment [A]:} In Subfigure~\ref{fig:final_zoom_500ms}, we observe that the GPU current varies between nearly 0\,A and around 25--30\,A in rapid succession. Such wide swings within a fraction of a second highlight the limitations of small-signal control assumptions and underscore the need for robust large-signal design approaches.

\begin{figure*}[t]
    \centering

    \begin{subfigure}[b]{0.48\textwidth}
        \centering
        \includegraphics[width=\textwidth]{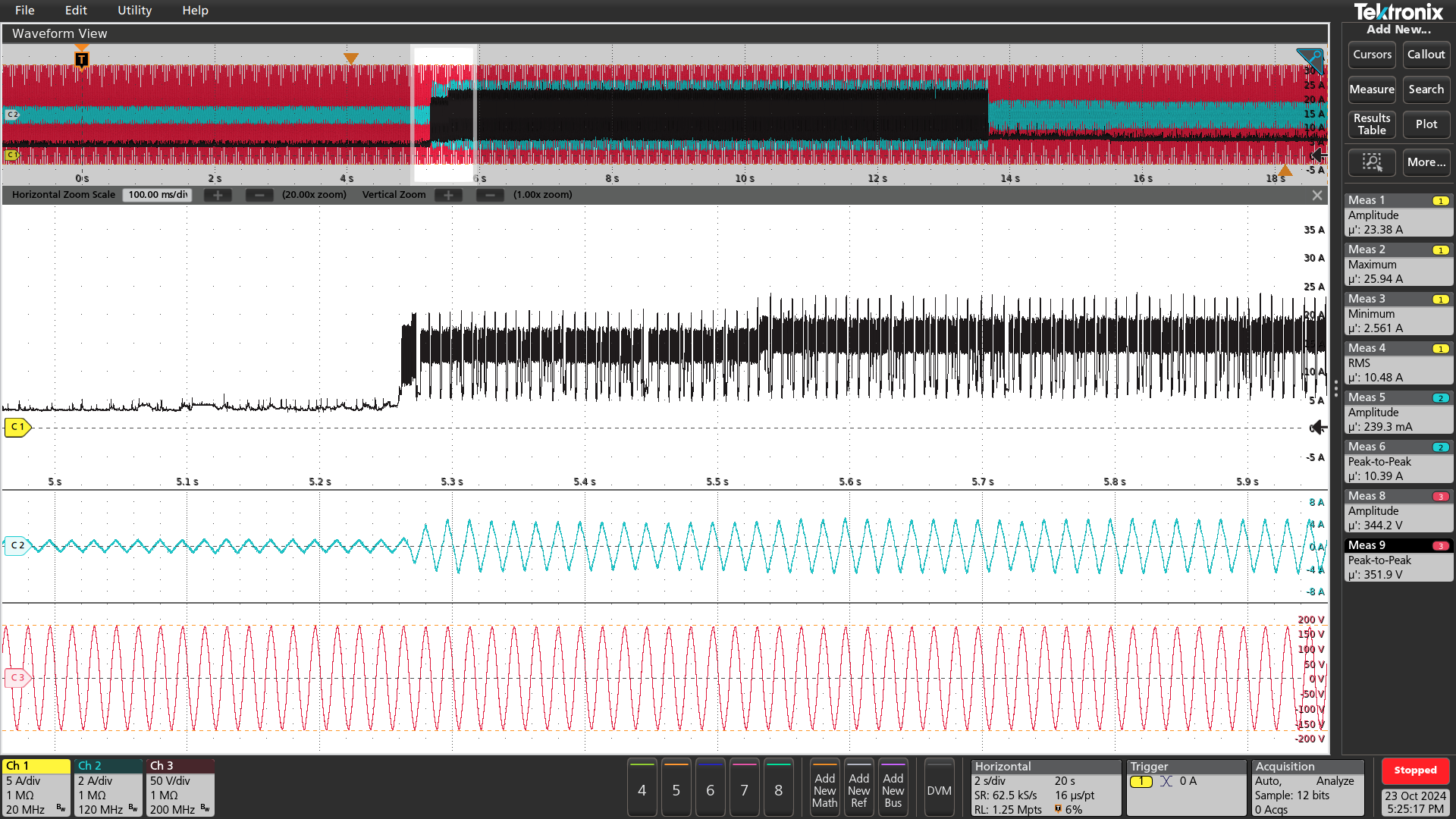}
        \caption{Initial capture showing the inference load-up transition for an RTX-4090 
        running a LLaMA-3.1 8B model. The GPU current (black trace) rapidly increases, 
        indicating the onset of inference operations.}
        \label{fig:inference_loadup}
    \end{subfigure}
    \hfill
    \begin{subfigure}[b]{0.48\textwidth}
        \centering
        \includegraphics[width=\textwidth]{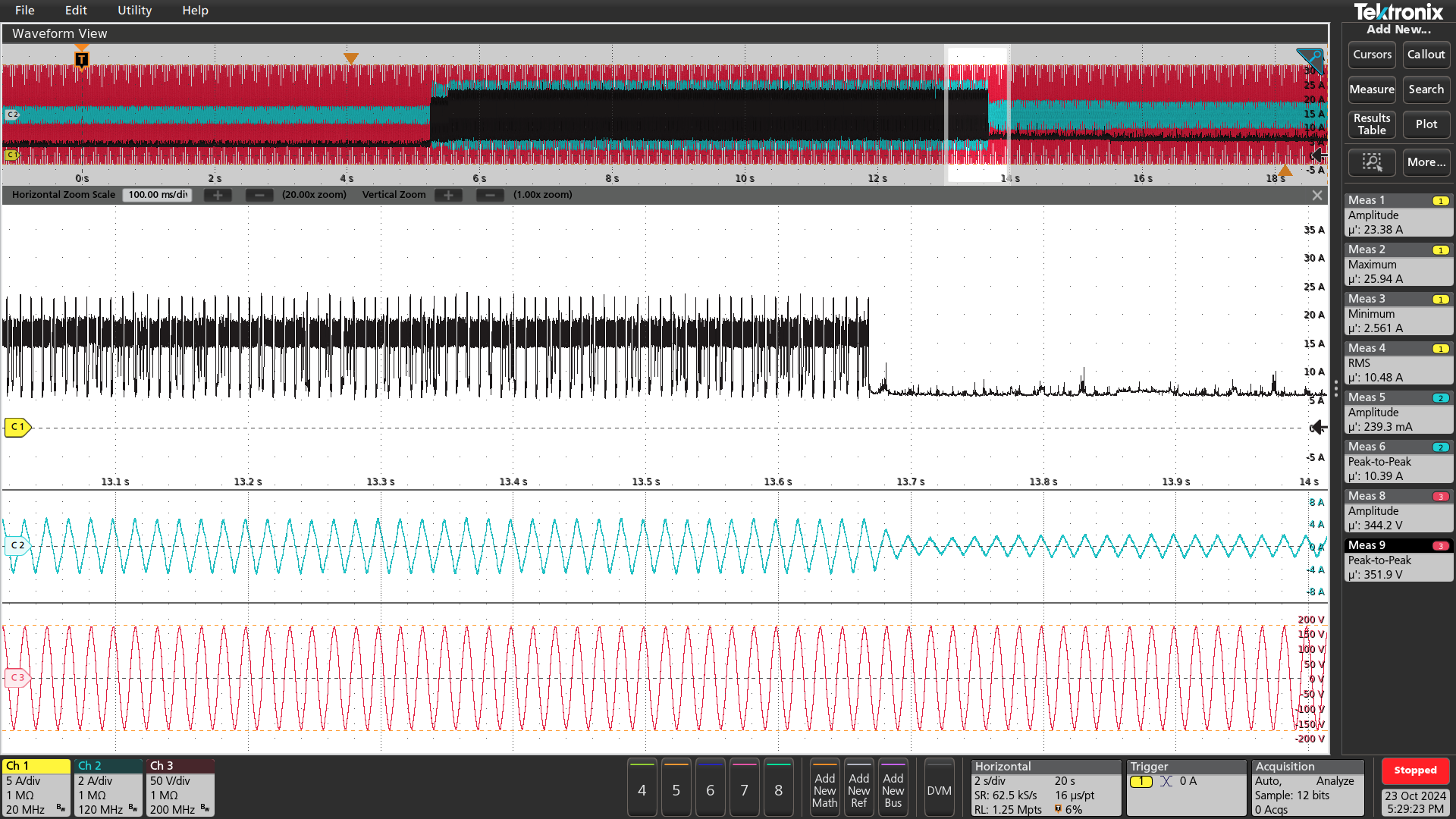}
        \caption{Subsequent capture illustrating the initial portion 
        of the load-down event (GPU returning to a lower-power state). 
        Note the abrupt decline in GPU current and the near-sinusoidal PSU input current.}
        \label{fig:inference_loaddown_1}
    \end{subfigure}

    \vspace{1em}
    
    \begin{subfigure}[b]{0.48\textwidth}
        \centering
        \includegraphics[width=\textwidth]{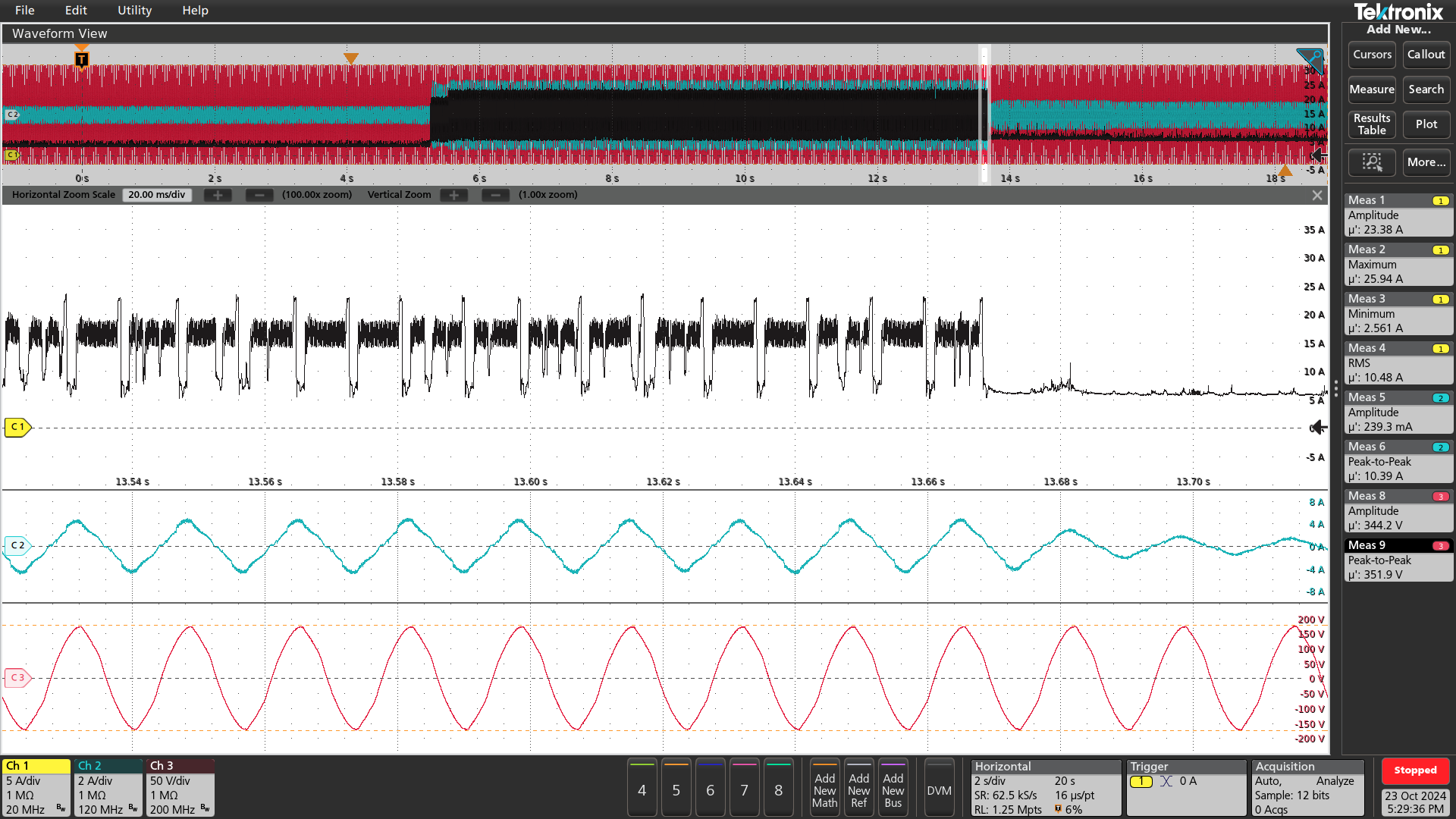}
        \caption{More detailed zoom of the negative transient, showing repeated 
        downward spikes in GPU current as the inference request load diminishes.}
        \label{fig:inference_loaddown_2}
    \end{subfigure}
    \hfill
    \begin{subfigure}[b]{0.48\textwidth}
        \centering
        \includegraphics[width=\textwidth]{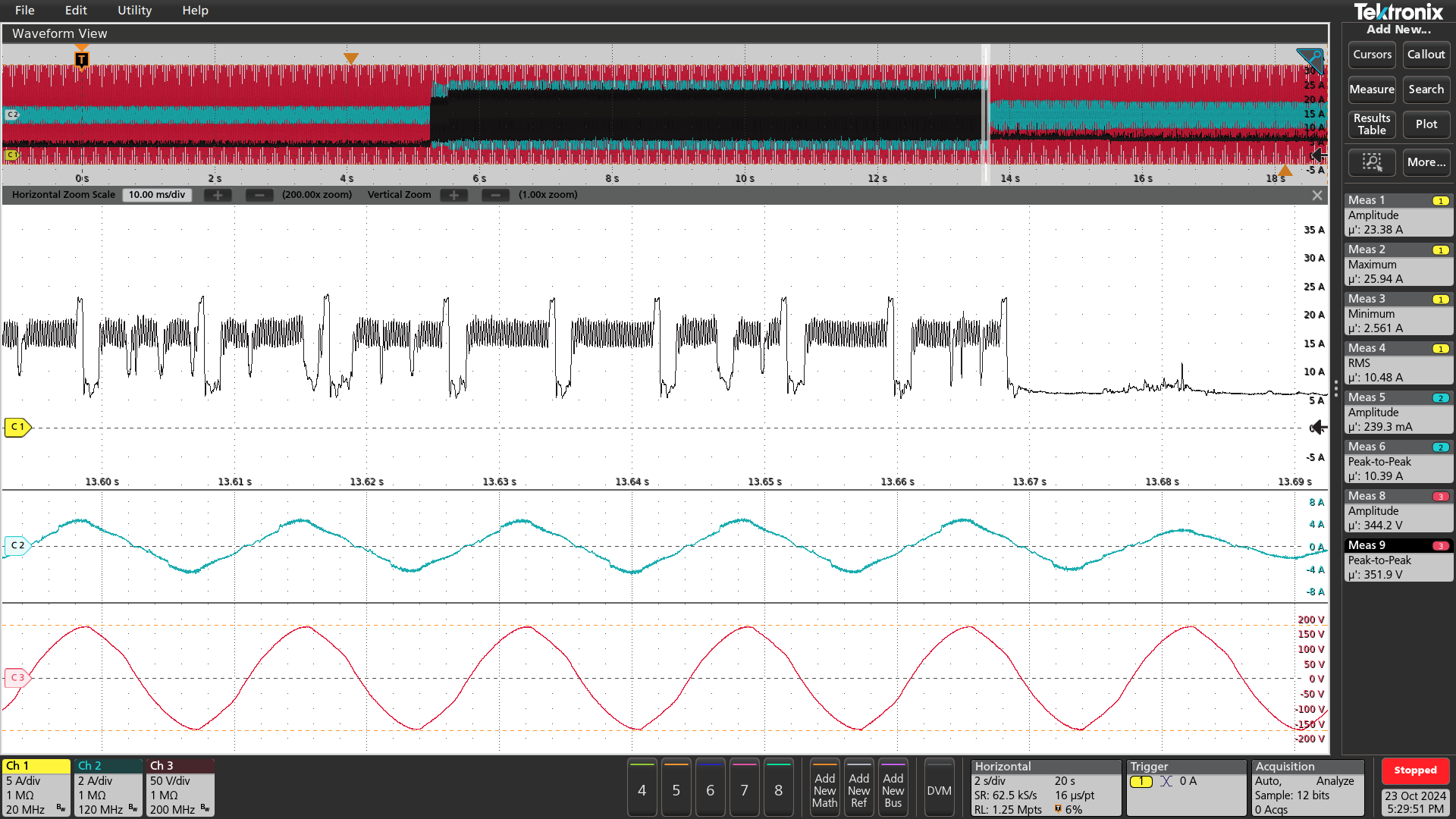}
        \caption{Millisecond-scale view capturing the final ramp to idle power. 
        The high-frequency ripples on the GPU current trace illustrate rapid 
        control-loop adjustments before settling.}
        \label{fig:inference_loaddown_3}
    \end{subfigure}

    \caption{Waveforms recorded during inference operations on an RTX-4090 running 
    a LLaMA-3.1 8B model. (a) The GPU load-up event draws substantial current 
    as the inference request begins, while (b)--(d) document the load-down stages 
    with progressively closer zoom, highlighting the abrupt negative transients 
    and the PSU’s response in stabilizing the supply voltage and current. Waveform from top to bottom: current of the GPU mother board, PSU current, PSU voltage.}
    \label{fig:inference_waveforms}
\end{figure*}

\paragraph{During LLaMA-3.1 8B Inference}
Figure~\ref{fig:inference_waveforms} shows how inference-driven load transitions differ from training checkpoints, yet still generate rapid current transients on the same RTX-4090 GPU. Subfigure~\ref{fig:inference_loadup} captures the moment the model begins processing a new inference request, ramping the current from a baseline level to roughly 20--25\,A in under 200\,ms. This spike in load indicates the shift from idle/standby operations to active compute.

Subfigures~\ref{fig:inference_loaddown_1}--\ref{fig:inference_loaddown_2} illustrate the load-down sequence, with repeated downward spikes in GPU current leading to near-idle operation. Meanwhile, the PSU input voltage (Ch3, red trace) remains relatively stable, confirming that local capacitors and PSU control loops effectively smooth some of the fastest edges. However, the abrupt negative transients still pose overvoltage risks if not properly managed—mirroring the training checkpoint concerns.

\noindent
\textbf{Comment [B]:} As discussed further in Section~\ref{sec:power_chain_dynamics}, handling these rapid load drops requires energy storage solutions or bi-directional converter topologies that can absorb or quickly curtail input power. The millisecond-level zoom in Subfigure~\ref{fig:inference_loaddown_3} highlights that at least a few AC cycles are involved in stabilizing current after the GPU load drop, indicating a need for carefully tuned control loops.

\subsection{Understanding AI Workload Power Consumption}

At their core, modern AI workloads, particularly LLMs, are built upon fundamental operations including matrix multiplications, attention mechanisms, and data movement operations. When an LLM processes information, it performs successive matrix transformations and self-attention computations across multiple layers, each contributing to the overall power demand. These operations create a characteristic power consumption pattern that can be decomposed into three primary components:
\begin{equation}
P_{op}(t) = P_{compute}(t) + P_{memory}(t) + P_{transfer}(t).
\end{equation}

The compute component, $P_{compute}(t)$, represents power consumed by matrix operations and attention computations, which can demand up to 90\% of peak power during intensive calculations. Modern LLMs implement these computations through specialized tensor processing units that perform highly parallel operations. The memory component, $P_{memory}(t)$, captures the power required for accessing model weights and activations. With current LLM architectures exceeding hundreds of billions of parameters, this component becomes increasingly significant. The transfer component, $P_{transfer}(t)$, accounts for data movement power, particularly critical in distributed training scenarios where model parallelism necessitates substantial communication between GPUs.

\noindent
\textbf{Comment [C]:} The combined effect of $P_{compute}(t)$ and $P_{memory}(t)$ frequently manifests as rapid load transients in Figures~\ref{fig:gpu_transient_zoom} and \ref{fig:inference_waveforms}. When large parameter blocks are read/written during an attention step or inference request, memory power surges coincide with intense compute kernels, further amplifying peak demands.

\subsection{Training Workload Dynamics}

Training operations represent one of the most demanding scenarios for power delivery systems. During these operations, the aggregate power consumption follows a characteristic pattern:
\begin{equation}
P_{train}(t) = P_{base} + \sum_{i=1}^{N} \alpha_i f_i(t) + \epsilon(t),
\end{equation}
where the baseline power $P_{base}$ often hovers around 60--70\% of peak load to maintain readiness for the next compute cycle. The summation term captures the power variations from different training phases, with $\alpha_i$ reflecting each phase’s intensity and $f_i(t)$ describing its duration or time-varying profile. The stochastic term $\epsilon(t)$ accounts for minor fluctuations due to micro-batching or asynchronous background tasks.

A typical training session proceeds through several distinct phases, each with unique power characteristics. During forward propagation, the accelerator maintains high but relatively steady power consumption as it processes data through the neural network layers. The backward propagation phase introduces more variability, as error gradients are computed and memory reads/writes intensify. This can lead to short bursts of full-load operation and subsequent partial idling between optimization steps. 
Together, these behavior patterns can induce the “sawtooth” or “spiky” load profiles evident in Figure~\ref{fig:gpu_transient_zoom}.

\vspace{0.8em}
\noindent
\textbf{Comment [D]:}  
It is important to distinguish between this “baseline” during a training loop—where the GPU remains at elevated clocks to rapidly process the next batch—and a \emph{true idle} state when no training is running. In many HPC environments, the “idle” portion of a training cycle still draws a substantial fraction of peak power (often 60--70\%), precisely because the hardware does not downclock fully between mini-batches. By contrast, a truly idle GPU (with no active training or inference tasks) may draw significantly less power, potentially nearer to 5--10\,A on the motherboard rail, depending on the GPU’s internal power management.  

\subsection{Inference Workload Characteristics}

Inference workloads present a different set of challenges, particularly in production environments handling multiple simultaneous requests. The power consumption during inference can be characterized as:
\begin{equation}
P_{inf}(t) = P_{idle} + \sum_{k=1}^{M} P_k(t) \cdot R_k(t),
\end{equation}
where $P_{idle}$ represents the baseline draw, and each term $P_k(t)\cdot R_k(t)$ models a specific request type’s power demand and arrival pattern.

\noindent
\textbf{Comment [E]:} In many inference scenarios, requests arrive in rapid bursts, causing GPU load to oscillate between idle (or low power) and near-peak consumption. Figures~\ref{fig:inference_loaddown_2}--\ref{fig:inference_loaddown_3} show how quickly the system returns to idle once a request completes, with negative transients that may reach 80--90\% of peak current in under one second.

\vspace{0.6em}
\noindent
\textbf{Comment [F]:} 
Closer inspection of our inference waveforms reveals that the baseline current may hover around 5\,A and can briefly approach 10\,A. This baseline likely stems from the GPU maintaining elevated clock states or memory readiness, even when nominally “idle.” In practice, one sees repeated short bursts from 5--10\,A up to 25--30\,A, followed by abrupt drops that occur within tens of milliseconds. The power supply’s internal buffering ensures the input current (Ch2) and AC voltage (Ch3) remain relatively stable, but on the motherboard rail (Ch1), these transitions can appear as rapid spikes and dips that challenge traditional control assumptions.

While individual inference operations may consume less power than a training step, latency demands can trigger frequent transitions. Batching can enhance efficiency by processing multiple requests simultaneously, but it can also alter the load profile by concentrating short, high-power bursts. This tradeoff between power efficiency and response time is central to practical inference deployments.

\subsection{System-Level Impact}

These workload characteristics create several fundamental challenges for power delivery system design. First, they require energy storage systems that can operate across multiple timescales. Local capacitors must handle microsecond-level transitions, while larger energy storage elements manage longer-term variations. The distribution of this energy storage through the system becomes a critical design consideration.

Second, control systems must manage power delivery across multiple timescales while maintaining stability. Traditional control approaches designed for slower-varying loads may not adequately handle the rapid transitions characteristic of AI workloads. This has driven the development of more sophisticated control strategies that can anticipate and respond to rapid power demand changes.

Third, protection systems must balance rapid response to potential faults with immunity to normal operation transients. The fast power transitions in AI workloads can look similar to fault conditions to traditional protection systems, requiring more sophisticated detection and coordination approaches.

\noindent
\textbf{Comment [G]:} In multi-megawatt HPC or AI clusters, aggregator effects can amplify transient peaks when dozens or hundreds of GPUs checkpoint or load/unload tasks nearly in unison. System-level designs must consider phase-shift or staggered scheduling approaches—especially when synchronizing gradient updates—to avoid large-scale power surges that exceed upstream supply capabilities.

These characteristics fundamentally influence power delivery system design decisions, from the selection of power converter topologies to the implementation of control and protection strategies. Successfully supporting AI workloads requires careful consideration of these patterns and their implications across all operational modes, as will be further examined in Section~\ref{sec:power_chain_dynamics}.

\section{Power Chain Architecture}
\label{sec:power_chain_arch}

The power chain architecture of modern data center infrastructures typically involves multiple cascaded conversion stages, each optimized for dynamic response and tuned to achieve the necessary balance between efficiency, reliability, and scalability. Such architectures carefully integrate converter topologies selected based on prevailing voltage levels, power ratings, switching frequency constraints, desired availability, and compliance with regional standards. Traditional multi-stage implementations have long been established as reliable and scalable solutions in a range of data center environments. These conventional approaches are grounded in well-understood practices and demonstrate consistently high performance. Meanwhile, emerging power chain architectures, which often seek to streamline conversion steps and improve system-wide efficiency, are increasingly gaining attention. Yet, their broader adoption remains subject to a variety of implementation constraints, operational complexities, and local regulatory considerations that must be addressed to ensure dependable performance over a data center’s lifespan.

\subsection{AC-based Power Chain Architecture}

\begin{figure*}[h]
  \centering
  \includegraphics[width=0.7\textwidth]{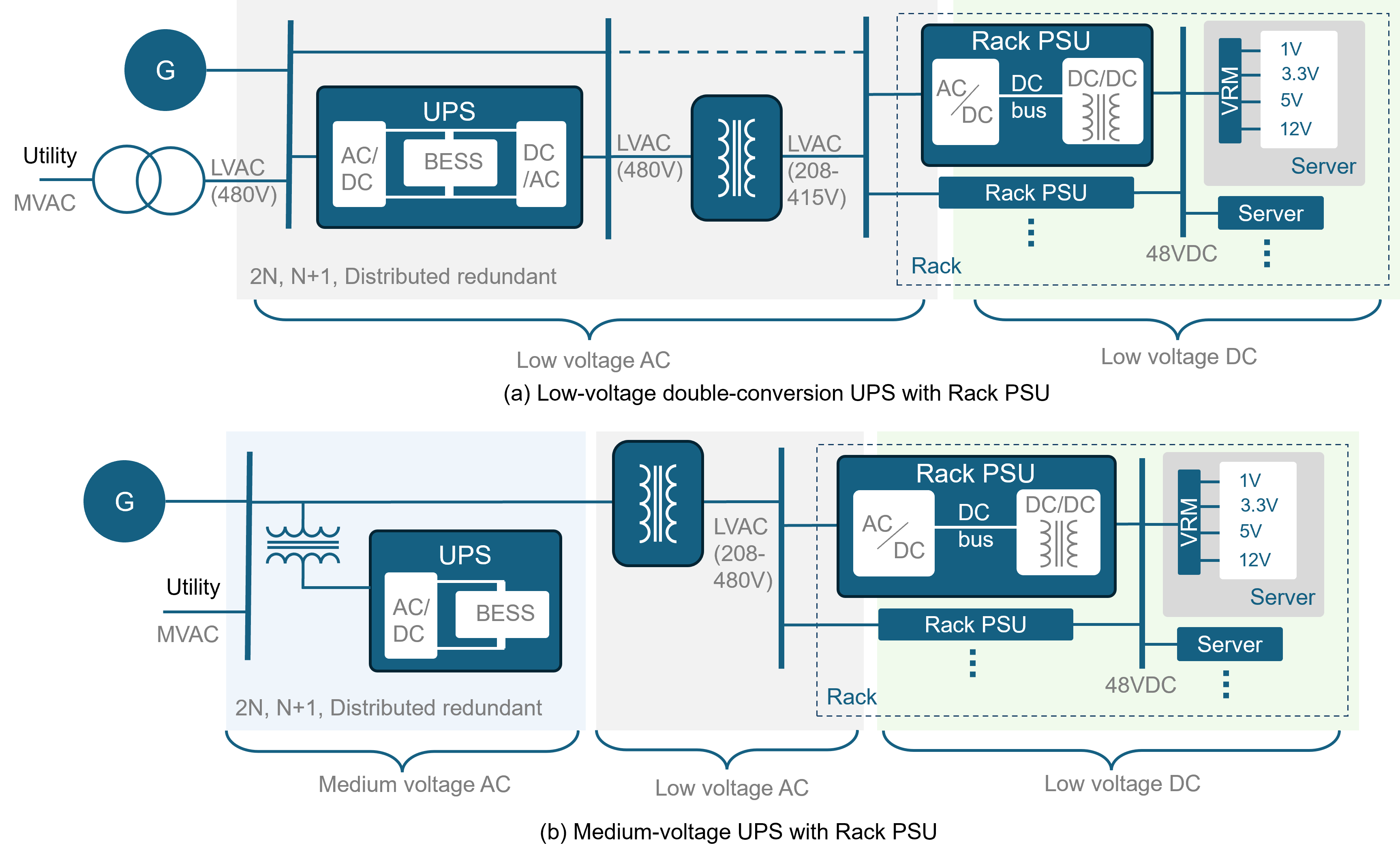}
  \caption{Power chain of GPU clusters (cooling excluded). The illustration shows multiple conversion stages (e.g., MVAC to LVAC, LVAC to DC, and final DC/DC stages) along with potential UPS integration.}
  \label{fig:power_chain_ups}
\end{figure*}

As depicted in Fig.~\ref{fig:power_chain_ups}(a), an LVAC-based data center typically steps down from MVAC (13.8\,kV, for example) to 480\,V or 415\,V AC, then uses a Uninterruptible Power Supply (UPS) stage before distributing to the rack-level PSUs.

The Low-Voltage AC (LVAC)-based architecture remains the predominant solution in contemporary data center designs, particularly in North American installations, due to historical infrastructural inertia and well-established engineering frameworks. This topology employs multiple cascaded conversion stages, each carefully tailored to address specific regional grid characteristics and operational demands. Since the efficiency of these conversion stages is closely tied to regional voltage standards and component design parameters, careful evaluation of the entire conversion cascade is necessary. For example, the initial transformation from medium-voltage AC (MVAC) down to a 480VAC distribution level typically achieves efficiencies between approximately 98.5\% and 99.2\%, depending on transformer quality and loading conditions.

In North America, the presence of additional transformation stages (e.g., from 480V to 277V) can introduce incremental efficiency penalties of roughly 1-2\% per stage. By contrast, many European implementations, which commonly use 400V/230V distributions, tend to have fewer intermediate steps, leading to slightly higher end-to-end efficiencies. In Asian deployments, such as those in Japan where 200V/100V standards dominate, unique challenges arise from more frequent transformation stages and from managing distribution losses effectively. These regional variations demand meticulous engineering approaches to optimize each stage’s design and achieve the best balance between efficiency, reliability, and compliance.

Further downstream, the Uninterruptible Power Supply (UPS) stage incorporates multiple conversion processes. Modern double-conversion UPS systems typically attain efficiencies around 94-96\% at rated loads, while line-interactive topologies may reach 97-98\% at full load, though at some expense in power conditioning capability. In practice, data centers may operate at only 30-50\% of rated capacity, meaning the actual in-service efficiency often deviates from these ideal figures. Hence, careful load profiling and region-specific design adjustments—such as right-sizing equipment to local standards—are critical to achieve intended performance targets.

In Fig.~\ref{fig:power_chain_ups}, one can also see the conceptual move to MVAC distribution (middle block) before stepping down to LVAC. This approach reduces cable losses but introduces different protection and compliance requirements \cite{hiperguard_mv_ups}.

Medium-Voltage AC (MVAC)-based power chain architectures represent a targeted evolutionary step that aims to reduce the number of conversion stages and thereby improve overall energy efficiency. They are often most effectively implemented in greenfield deployments, where the entire infrastructure can be designed holistically to accommodate higher distribution voltages. Nonetheless, regional standards play a significant role in influencing the feasibility of these systems. Voltage regulation requirements, for instance, vary widely: in North America, a ±5\% tolerance is common, while some European jurisdictions permit ±10\% or other region-specific ranges. Similarly, the management of fault currents becomes more critical at higher voltages, necessitating careful device selection and coordination of protection schemes that align with local safety and reliability standards.

Power quality metrics must also be carefully considered. In North America, IEEE Standard 519-2014 generally stipulates a Total Harmonic Distortion (THD) limit of less than 5\% at the point of common coupling (PCC). European installations often refer to IEC 61000-3-2 or related standards, which may impose stricter harmonic distortion limits, depending on equipment class and load type. Additionally, regional differences in grid stability, power factor requirements (often in the 0.95-0.98 leading range), and local code compliance significantly influence the ultimate design of MVAC-based architectures. Addressing these region-specific conditions is essential for ensuring that the chosen topology offers tangible reliability and efficiency improvements.

\subsection{DC-based Power Chain Architecture}

\begin{figure}[!t]
  \centering
  \includegraphics[width=\columnwidth]{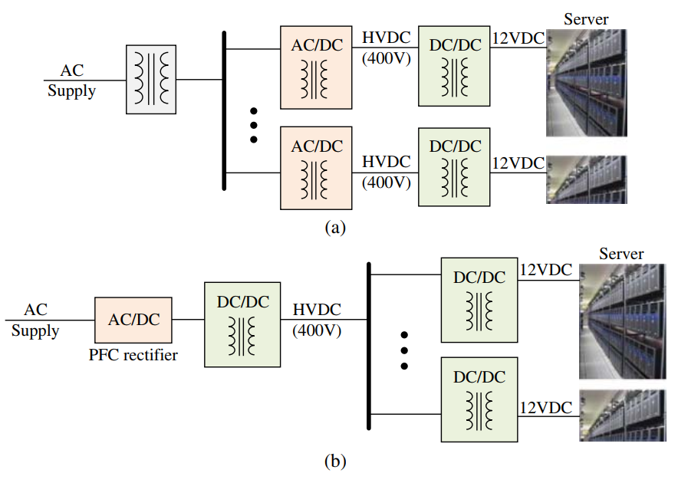}
  \caption{Data center power delivery system with 400V DC bus.}
  \label{fig:400Vdc}
\end{figure}

DC-based architectures introduce the possibility of more direct and streamlined energy delivery paths by eliminating certain AC-DC and DC-AC conversion stages (Fig.~\ref{fig:power_chain_ups}(b)). Researchers have evaluated 400\,V DC distribution in telco and data centers, finding notable efficiency improvements \cite{4448733,quantitative_comparison_ac_dc}. Google’s 400\,V DC rack proposal likewise addresses ML/AI applications \cite{GoogleOCP2024}, conceptually showing in Fig.~\ref{fig:400Vdc}. In practice, the realized gains may be more modest once all practical factors are considered, including the availability of DC-rated equipment, the complexity of grounding schemes, and the intricacies of safety compliance across different markets.

Grounding requirements, for example, are highly region-dependent and can vary from redundant ground paths mandated in North America to more permissive floating DC system configurations permitted under specific conditions in certain European installations. Asian markets, with their diverse regulatory landscapes, often adopt mixed approaches to grounding and protection. Voltage level selection must take into account the required isolation margins—often in the range of 1.5 to 2.5 times the nominal voltage—as well as semiconductor device ratings and the availability of DC-rated equipment that satisfies local regulatory frameworks.

Protection strategies for DC systems are often more challenging due to the lack of inherent zero-crossing in DC fault currents. Arc flash energy levels must be carefully managed, and the limited availability of high-current DC breakers in some markets can complicate device selection and compliance. Ground fault detection thresholds also vary, with acceptable sensitivities ranging from as low as 3mA up to 30mA, depending on local standards and operational practice.

Although a two-stage power chain (e.g., direct 48\,V DC architecture) can theoretically reduce the number of conversion steps and improve efficiency, its practical use in hyperscale (100MW+) data centers remains to be verified. The high currents and significant distribution distances at such scales create challenges in terms of voltage drop, cable sizing, and equipment availability. Nevertheless, smaller or specialized deployments (e.g., telecom or edge sites) may benefit from a two-stage approach due to shorter distribution paths and lower overall power demands.


\section{Power Chain Dynamics}
\label{sec:power_chain_dynamics}

\noindent
In this paper, we define the “final stage” as the power interface to the \emph{utility or generator}, while the “first stage” refers to the \emph{rack-level} or \emph{GPU-level} power modules. 
This labeling follows an industrial perspective: large upstream converters (e.g., MVAC--LVAC or UPS front-ends) often have lower control bandwidths and higher power ratings, making them a global bottleneck for rapid transients. 
Downstream Voltage Regulator Modules (VRMs) near the GPUs may switch at tens or hundreds of kHz, but their fast local loops can only buffer small energy perturbations in the millisecond/microsecond range.

Furthermore, a \textbf{load jump} is comparatively straightforward to handle because a data center can orchestrate a pre-charge or ramp-up sequence. 
A \textbf{load drop}, on the other hand, can occur suddenly (e.g., if a GPU crash stops computation), leaving the power chain with excess energy that must be safely absorbed or redirected. 
Such negative transients demand robust design considerations in the upstream (final-stage) converter, including bi-directional energy paths or advanced control schemes. High-rate energy storage, such as supercapacitors, is particularly effective at absorbing sudden negative transients \cite{lee2010ultracapacitor}.

\medskip

\subsection{Practical PSU Topologies and Control}
\label{subsec:psu_topo_control}

\begin{figure}[ht]
  \centering
  \includegraphics[width=\columnwidth]{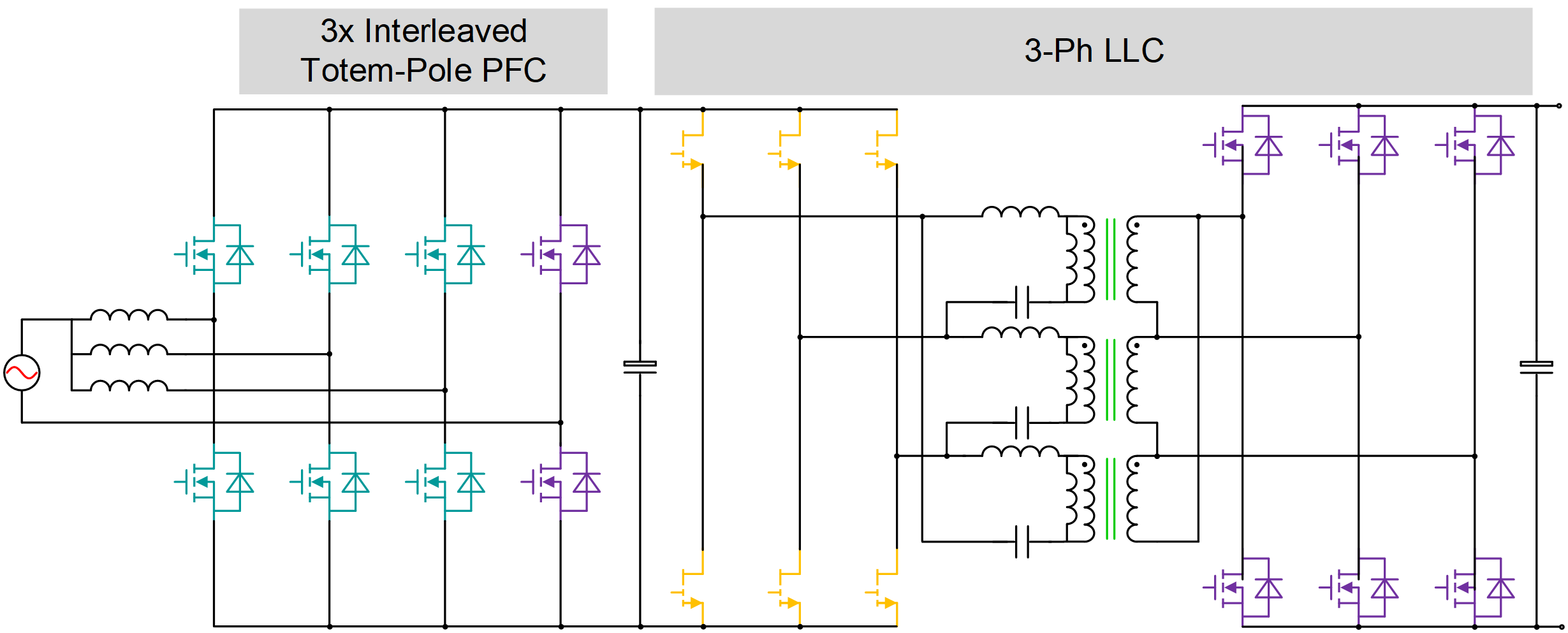}
  \caption{An 8.5 kW PSU architecture from Navitas, illustrating a Totem-Pole PFC front-end and a DC/DC stage.}
  \label{fig:8.5kW}
\end{figure}

\begin{figure}[ht]
  \centering
  \includegraphics[width=\columnwidth]{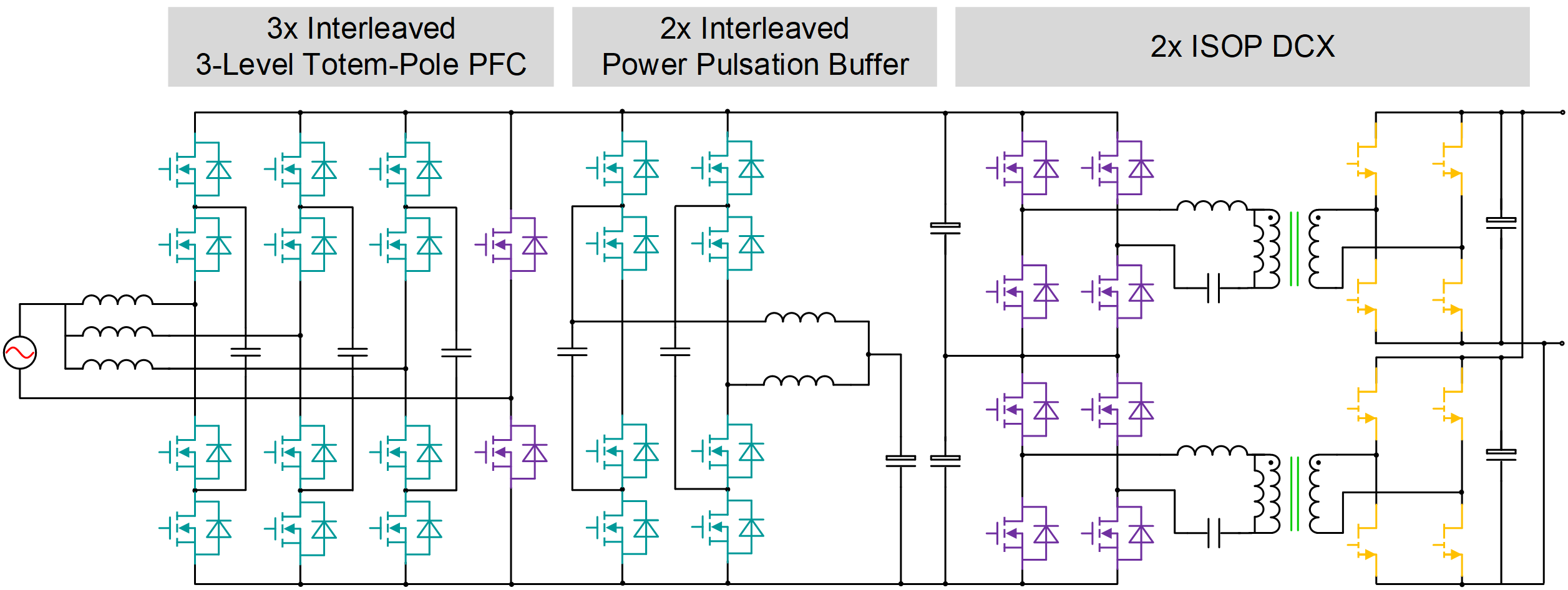}
  \caption{A 12 kW PSU from Infineon, featuring advanced interleaved PFC and DC/DC topologies designed for high-density power conversion.}
  \label{fig:12kW}
\end{figure}

\paragraph{Totem-Pole Power Factor Correction (PFC) and Interleaved PFC Examples}
Figure~\ref{fig:8.5kW} illustrates an 8.5\,kW power supply that employs a \emph{totem-pole PFC} topology on the AC front end, replacing the traditional diode bridge with a bidirectional switch leg (often using GaN or SiC devices) to reduce conduction losses \cite{navitas_2024}. Among its key design features is \textbf{Reduced Bridge Losses}, achieved by using an active leg in place of a full diode-bridge rectifier. This minimizes the number of forward-voltage drops and thereby boosts overall efficiency, which is especially vital at multi-kilowatt scales. Another important aspect is the \textbf{High Switching Frequency} made possible by fast-switching GaN or SiC MOSFETs (often denoted $S_1$ and $S_2$). Operating at higher frequencies decreases the size of magnetic components while still maintaining high efficiency. In addition, the totem-pole stage provides \textbf{Bidirectional Capability}, allowing it to be designed for partial energy return (limited by control) during negative transients such as load drops, so that some energy can flow back to the source or be dissipated in a controlled manner.

Figure~\ref{fig:12kW} highlights a 12\,kW PSU with \emph{interleaved PFC} stages and a subsequent DC/DC converter \cite{infineon_oktobertech24}. By splitting the input current into multiple parallel phases, this interleaving scheme provides \textbf{Current Sharing}, which reduces the RMS current stress in each inductor and switch, permitting the use of smaller, more efficient components. Furthermore, phase interleaving achieves \textbf{Reduced Ripple} on both the AC and DC sides, leading to higher overall efficiency and lower Electromagnetic Interference (EMI). It also enables \textbf{Scalability}, a crucial consideration as data-center rack powers rise into the tens of kilowatts per PSU. Interleaving is thus widely adopted in industry to increase power density while maintaining manageable thermal designs. Downstream of the PFC, the DC/DC stage often relies on resonant or soft-switching methods (e.g., LLC resonant or phase-shift full-bridge) to further improve efficiency under moderate transients.

\medskip

\begin{figure}[ht]
  \centering
  \includegraphics[width=0.5\columnwidth]{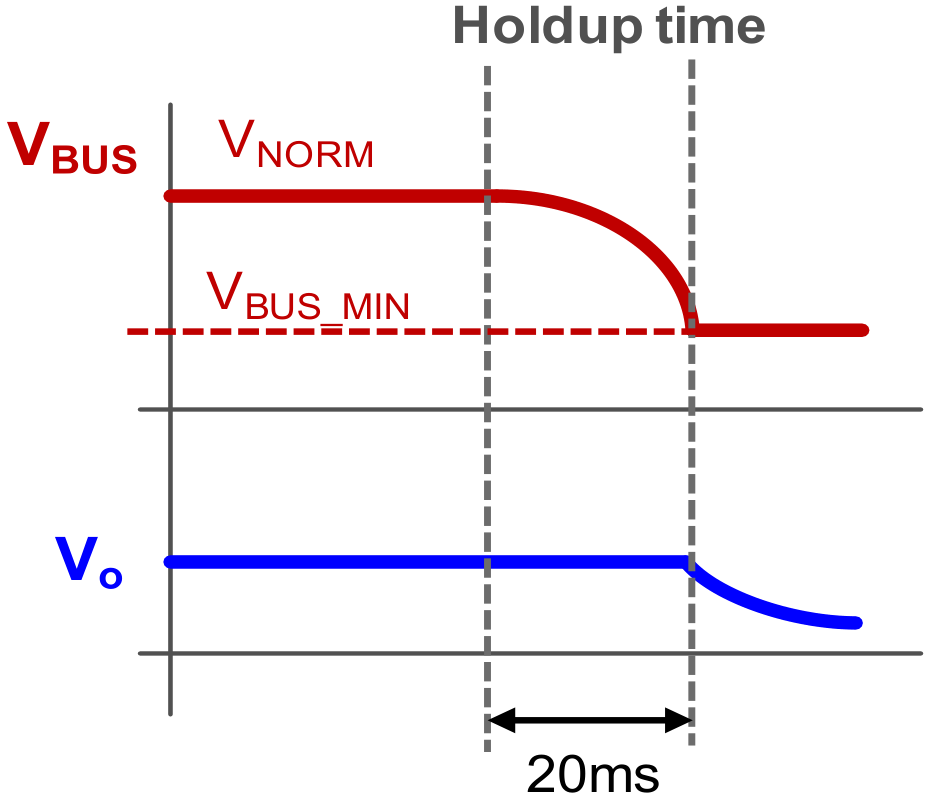}
  \caption{Hold-up time operation of PFCs, showing how DC-link capacitors buffer short-term outages.
}
  \label{fig:holdup}
\end{figure}

\paragraph{Dual-Loop Control and Hold-Up Time}
In a typical \emph{dual-loop control} scheme for the PFC front end, the \textbf{Outer Voltage Loop} maintains the desired DC bus voltage ($V_{\mathrm{BUS}}$) by generating a reference current, $i_\mathrm{ref}$, proportional to the load demand. In parallel, the \textbf{Inner Current Loop} modulates the PWM duty cycle to shape the inductor current, ensuring it follows $i_\mathrm{ref}$ and remains sinusoidal while staying in phase with the AC input. Feedforward paths (e.g., measuring $v_\mathrm{in}$ or predictive load monitoring) can be used to mitigate large overshoots or undershoots before the feedback loops fully react. 
\textit{For example, in high-power GPU clusters, the controller may incorporate predictions of upcoming load steps and adjust the reference preemptively.}

Figure~\ref{fig:holdup} shows the importance of \emph{hold-up time}, where a DC-link capacitor buffers the supply if there is an abrupt input failure. The capacitor keeps the \textbf{Bus Voltage} ($V_{\mathrm{BUS}}$) near nominal until it discharges, and a \textbf{Minimum Acceptable Voltage} ($V_{\mathrm{BUS, MIN}}$) ensures downstream regulators still operate. The margin between $V_{\mathrm{NORM}}$ and $V_{\mathrm{BUS, MIN}}$ determines how much energy is stored for bridging short outages (e.g., $t_{\mathrm{hold}} \approx 20$\,ms). Together, these control and design features ensure rapid handling of \emph{load jumps} and safer absorption of \emph{load drops}, which is particularly vital in high-power data-center applications where sudden GPU load changes occur on millisecond or even microsecond timescales.

\subsection{Small-Signal vs.\ Large-Signal Dynamics}

Traditionally, data center power designs rely on small-signal models linearized around a nominal operating point (e.g., 50--80\% load). For modest fluctuations, this approach suffices for stability checks and controller tuning. However, \textbf{AI workloads} can exhibit large load swings (20\% $\to$ 100\% $\to$ idle) on millisecond timescales, making small-signal assumptions inaccurate because the operating point is no longer “close” to nominal. In industrial practice, Bode plots and phase-margin analyses remain useful for local-loop tuning, but they do not capture the full nonlinear or large-signal behavior that arises during abrupt GPU load changes. Consequently, system architects have begun to incorporate time-domain or piecewise-linear analyses to better account for these rapid and sizable transients.

\subsubsection{Representative Small-Signal Transfer Function}

In the small-signal domain, each AC/DC or DC/DC stage can be approximated by:
\begin{equation}
    G_i(s) \;\approx\; \frac{V_{o,i}(s)}{I_{o,i}(s)}
    \;=\; \frac{k_i}{\bigl(1 + \tfrac{s}{\omega_{z,i}}\bigr)\,\bigl(1 + \tfrac{s}{\omega_{p,i}}\bigr)},
\end{equation}
where \(V_{o,i}\) and \(I_{o,i}\) are deviations around the operating point, and \(\omega_{z,i}\) and \(\omega_{p,i}\) are the zero and pole frequencies, respectively. When multiple stages are cascaded, the overall transfer function becomes
\begin{equation}
    G_{\mathrm{system}}(s) \;=\; \prod_{i=1}^{N} G_i(s),
\end{equation}
thus combining the frequency responses of each power stage. In many data center environments, the dominant (slowest) pole is often located in an upstream converter, restricting the overall bandwidth to only a few kilohertz—even if downstream VRMs can switch in the hundreds of kilohertz range or even higher. As a result, large-signal performance often ends up limited by the slower, high-power front-end units rather than by the fast local loops near the load.

\subsection{Large-Signal Modeling and Controller Structures}
\label{subsec:large_signal_controller_structures}

Modern data center power chains typically employ a hierarchical control strategy to handle both steady-state regulation and sudden transients. At the most granular level, an \emph{Inner Current Loop (ICL)} ensures that the converter’s input or output current accurately tracks a reference signal, providing rapid fault protection and power factor control when working with AC sources. Above that, an \emph{Outer Voltage Loop (OVL)} regulates the DC-link or bus voltage by adjusting the inner loop’s current reference to match variations in load demand. Ultimately, a higher-level \emph{Supervisory Control} system ties together multiple parallel modules, oversees battery or UPS operation, and ensures compliance with grid or generator constraints. This supervisory layer often communicates with facility management systems in real time, particularly in large-scale data centers.

Before focusing on hierarchical loop structures, it is instructive to present a comprehensive \emph{large-signal} model of a final-stage AC/DC converter rated at \(P_{\mathrm{rated}}\). For simplicity, consider a single-phase AC input at voltage \(v_{\mathrm{ac}}(t)\), a rectifier stage, and a controlled switch (or full-bridge) with an output filter inductor \(L\) feeding a DC-link capacitor \(C\). Let
\begin{itemize}
    \item \(i_{\mathrm{L}}(t)\) be the inductor current,
    \item \(v_{\mathrm{L}}(t)\) be the voltage across the inductor,
    \item \(v_{\mathrm{dc}}(t)\) be the DC-link voltage,
    \item \(d(t)\in[0,1]\) be the duty ratio,
    \item \(P_{\mathrm{load}}(t)\) be the instantaneous load power (e.g., a GPU load).
\end{itemize}

A simplified set of large-signal dynamic equations for this power stage can be written as:
\begin{equation}
    L \,\frac{d\,i_{\mathrm{L}}(t)}{dt} = v_{\mathrm{ac,rect}}(t) - \bigl[1-d(t)\bigr]\,v_{\mathrm{dc}}(t), \\
\end{equation}
\begin{equation}
    C \,\frac{d\,v_{\mathrm{dc}}(t)}{dt} = d(t)\,i_{\mathrm{L}}(t)\,v_{\mathrm{ac,rect}}(t) \;-\; P_{\mathrm{load}}(t),
\end{equation}
where \(v_{\mathrm{ac,rect}}(t)\) is the rectified AC input voltage. Note that this model assumes ideal switching and negligible losses for brevity; more sophisticated versions include conduction losses, switching losses, and any EMI filter dynamics. These \emph{large-signal} equations highlight how rapid changes in \(d(t)\) or \(P_{\mathrm{load}}(t)\) can directly impact both \(i_{\mathrm{L}}(t)\) and \(v_{\mathrm{dc}}(t)\). Such a state-space formulation is crucial for understanding \emph{transient} phenomena that small-signal linearizations may overlook.

One way to formalize the hierarchical loops acting on this converter is to note that the outer voltage loop calculates an error,
\begin{equation}
  e_v(t) = V_{\mathrm{dc,ref}} - V_{\mathrm{dc}}(t),
\end{equation}
and transforms this into a reference current \(i_{\mathrm{ref}}(t)\) via a compensator \(K_v(s)\). In turn, an inner current loop compares \(i_{\mathrm{in}}(t)\) with \(i_{\mathrm{ref}}(t)\), and a compensator \(K_i(s)\) generates the PWM duty ratio \(d(t)\). While small-signal models and frequency-domain analysis provide a baseline for loop design and stability, \emph{large-signal} events such as a GPU load swing from near-idle to full capacity can cause sudden changes in \(i_{\mathrm{ref}}\). Addressing these abrupt shifts demands robust energy buffering---through bulk DC-link capacitors or supercapacitors---and fast control action to maintain the DC voltage within safe limits.

During these large swings, the \emph{energy-balance perspective} often proves more accurate than linear approximations. In particular, changes in the DC-link voltage can be expressed as
\begin{equation}
    C \,\frac{d\,V_{\mathrm{dc}}(t)}{dt}
    = i_{\mathrm{in}}(t)\,V_{\mathrm{in}}(t) \;-\; P_{\mathrm{load}}(t),
\end{equation}
where \(C\) is the DC-link capacitance, \(i_{\mathrm{in}}(t)\,V_{\mathrm{in}}(t)\) is the input power, and \(P_{\mathrm{load}}(t)\) is the instantaneous load power. If \(P_{\mathrm{load}}\) falls drastically, e.g., due to a sudden GPU stop, the mismatch between input and load power causes \(V_{\mathrm{dc}}(t)\) to rise quickly. Hence, the converter must either ramp down \(i_{\mathrm{in}}\) or divert the surplus energy into a resistor bank, battery, or supercapacitor---failure to do so can lead to excessive voltage spikes.

\textbf{Nonlinear State-Space Representation.}  
Large-signal events can be analyzed more precisely by casting the above dynamic equations into a general \emph{nonlinear state-space} form. Let the state vector be 
\begin{equation}
x(t) = 
\begin{bmatrix}
    i_{\mathrm{L}}(t)\\ 
    v_{\mathrm{dc}}(t)
\end{bmatrix}, 
\quad
u(t) = 
\begin{bmatrix}
    d(t)\\
    P_{\mathrm{load}}(t)
\end{bmatrix}.
\end{equation}
Then we can write 
\(\dot{x}(t) = f(x(t), u(t))\) and \(y(t) = g(x(t),u(t))\), where \(y(t)\) might include measurable outputs such as \(v_{\mathrm{dc}}(t)\) or \(i_{\mathrm{L}}(t)\). This \emph{nonlinear} representation is particularly useful for analyzing operating points that deviate significantly from nominal conditions. It also provides the foundation for advanced \emph{nonlinear control} approaches (e.g., \emph{sliding-mode control}, \emph{feedback linearization}, or \emph{Lyapunov-based} methods) that can robustly handle abrupt changes in \(P_{\mathrm{load}}(t)\).

To ensure safe and efficient operation under large-signal disturbances, several control-theoretic investigations may be employed:

\begin{itemize}
    \item \emph{Phase-Plane Analysis:} By examining trajectories in the \((i_{\mathrm{L}}, v_{\mathrm{dc}})\) plane, one can determine how quickly the system recovers from transients. This is especially useful for identifying boundary conditions under extreme load steps.
    \item \emph{Lyapunov Stability Criteria:} Constructing a suitable Lyapunov function can confirm global stability of the converter system under specific controller parameters. This is valuable in data center environments where reliability is paramount.
    \item \emph{Sliding-Mode Control:} The discontinuous nature of large load steps can be mitigated by forcing the system to “slide” along a predefined manifold, offering robustness against parameter variations and sudden transients.
    \item \emph{Feedback Linearization:} By algebraically transforming the nonlinear system, feedback linearization methods can help achieve linear-like performance over a wide operating range. This approach is particularly powerful if precise knowledge of system parameters is available.
\end{itemize}

\textbf{Extended Energy-Balance Control.}  
Beyond simply regulating voltage or current setpoints, an \emph{extended energy-balance} control framework incorporates additional actions to manage the flow of surplus or deficit energy. For instance, when \(P_{\mathrm{load}}\) drops sharply, an \emph{adaptive feedforward} term can modulate \(K_i(s)\) or \(K_v(s)\) to quickly reduce the input current, thus minimizing overshoot in \(V_{\mathrm{dc}}(t)\). Conversely, when the load spikes rapidly, the feedforward signal can preemptively boost the converter’s current reference to mitigate under-voltage dips. This holistic viewpoint of energy transfer also allows for coordinated use of local storage elements (e.g., supercapacitors) that can further buffer large power swings.

\textbf{Predictive and Constraint-Based Strategies.}  
When large transients are frequent or have tight performance requirements, \emph{model predictive control (MPC)} offers a flexible solution. By numerically optimizing a cost function over a finite horizon, the controller can account for real-time system constraints (e.g., current and voltage limits) and compute the optimal duty cycle \(d(t)\). This predictive method is particularly effective at maintaining safe operating conditions under sudden GPU load ramps or drops, where classical linear compensators may struggle with rapid nonlinear dynamics. In addition, constraint-handling ensures that key limits are never violated, which is critical for protecting both the power electronics and the load.

\textbf{Coordinated Supervisory Actions.}  
On a higher level, supervisory controllers can integrate large-signal models into their decision-making. By continuously monitoring predicted load profiles, the supervisory layer can schedule power module activation or coordinate energy storage dispatch (e.g., batteries, supercapacitors) to absorb or supply transient power. In cases where multiple parallel converter modules operate in tandem, the controller can dynamically redistribute load among modules to spread out thermal and electrical stresses, enhancing both system resilience and efficiency. Moreover, advanced supervisory policies may anticipate large load steps by pre-charging energy storage elements, thereby reducing the magnitude of transients seen by downstream converters.

Overall, the \emph{large-signal} model and the associated control strategies introduced above underscore the complexity of powering modern data centers. While classic small-signal analyses remain crucial for nominal design and linearized stability checks, the \emph{nonlinear} nature of massive load swings requires complementary methods such as energy-balance principles, phase-plane investigations, and predictive optimization. Future work may focus on unifying these approaches into a cohesive framework that seamlessly handles both routine regulation and worst-case fault scenarios.

\subsection{Cascaded Power Stage Analysis and System Dominance}

Data center power typically flows through multiple stages, for instance: MVAC\(\to\)LVAC\(\to\)48\,V\(\to\)12\,V\(\to\)GPU. Each stage contains its own inductors, capacitors, and control loops, all of which influence how quickly or slowly power can ramp. The \textbf{final-stage} AC/DC converter, or UPS, often incorporates large magnetics to handle high power levels and thus cannot change its current as fast as smaller downstream regulators. Meanwhile, local VRMs at the GPU boards can respond in microseconds but have limited energy storage. Consequently, when a GPU suddenly demands an enormous current jump, the local capacitors absorb the immediate surge, while the upstream source gradually ramps its input current within the confines of its bandwidth. Prolonged heavy load then propagates upstream once the small buffers at the VRM stage are depleted.

The concept of \emph{overall bandwidth constraint} emerges from this cascade. Since each stage has a certain crossover frequency \(\omega_{c,i}\), the slowest stage effectively sets the pace for system-level transients:
\begin{equation}
    \omega_{\mathrm{cl}} \approx \min_{1 \,\le\, i \,\le\, N} \bigl(\omega_{c,i}\bigr).
\end{equation}
If the final-stage converter’s control bandwidth is much lower than that of the VRM, the entire data center power chain remains bottlenecked by the slower unit. In industrial data centers, where multiple MWs of power might be supplied by paralleled AC/DC modules, this constraint poses a practical limit on how quickly large load swings can be supported without risking system instability or voltage excursions.

\subsubsection{Dominant Power Stage Dynamics}

While downstream loops might settle in just a few microseconds, the outer voltage loop of the upstream converter or UPS may require hundreds of microseconds—or even milliseconds—to react fully to a large load jump. For instance, when a GPU load rises from 20\% to 100\% in a fraction of a millisecond, local bus capacitors have to supply the initial current surge. If the surge endures, the final stage begins ramping its AC input current. Conversely, if the load drops abruptly from 100\% to 20\% or even near-zero, the system faces a sudden surplus of energy in the DC link. Quickly redirecting or dissipating this surplus is a major challenge: otherwise, the DC voltage can spike, triggering protective shutdowns or damaging hardware.

In particular, negative transients pose serious risks when battery or UPS systems are designed mainly for outage protection, i.e., discharging rather than charging at high rates. A GPU crash or job completion might halt a load consuming hundreds of kilowatts, causing
\begin{equation}
    C \,\frac{d\,V_{\mathrm{dc}}}{dt} \;>\; 0,
\end{equation}
which leads to a voltage swell unless the incoming power is quickly curtailed. Active control can reduce \(i_{\mathrm{in}}\) at the AC/DC stage, while dissipative elements or bi-directional energy storage systems (BESS) can help safely absorb the excess. Failure to handle these negative transients can produce nuisance tripping of protective circuits or outright hardware failure.

\subsubsection{Transient Scenarios and Rapid Load Changes}

Large load swings typically manifest as either \emph{positive transients} (ramp-up) or \emph{negative transients} (ramp-down). In a positive transient, such as a GPU jumping from idle to near-TDP, local capacitors buffer the immediate surge, and the final stage ramps up gradually to replenish the DC-link voltage. Often, techniques like pre-charging or staged enable sequences are employed to limit inrush currents.

Negative transients are generally more critical. When the load drops abruptly—whether intentionally, through normal job completion, or unexpectedly, via a GPU fault—\(P_{\mathrm{load}}\) can plummet from a high value to near-zero. In some industrial systems, a predictive controller might use GPU job-scheduling signals to anticipate an impending load reduction, issuing commands to reduce \(i_{\mathrm{in}}\) or route current into a fast-charging storage element. Without these measures, the DC-link voltage can spike within microseconds or milliseconds, potentially leading to equipment damage.

Finally, the special case of a sudden load drop—where the GPU instantly ceases consumption—represents the worst-case scenario for many power designs. Unless the excess energy is dumped or stored in a matter of microseconds, the rapid rise of \(V_{\mathrm{dc}}(t)\) may exceed design tolerances. Therefore, robust negative-transient handling often includes resistor dump circuits, supercapacitor banks, or bi-directional batteries with high charge acceptance capability. Such solutions improve the overall reliability and stability of AI-oriented data centers by ensuring safe operation even under extreme load transitions.

\section{Case Study}

\subsection{PSU Inertia Case Study}
\label{subsec:psu_inertia}

\begin{figure*}[t]
  \centering
  \begin{subfigure}[b]{0.6\textwidth}
    \includegraphics[width=\textwidth]{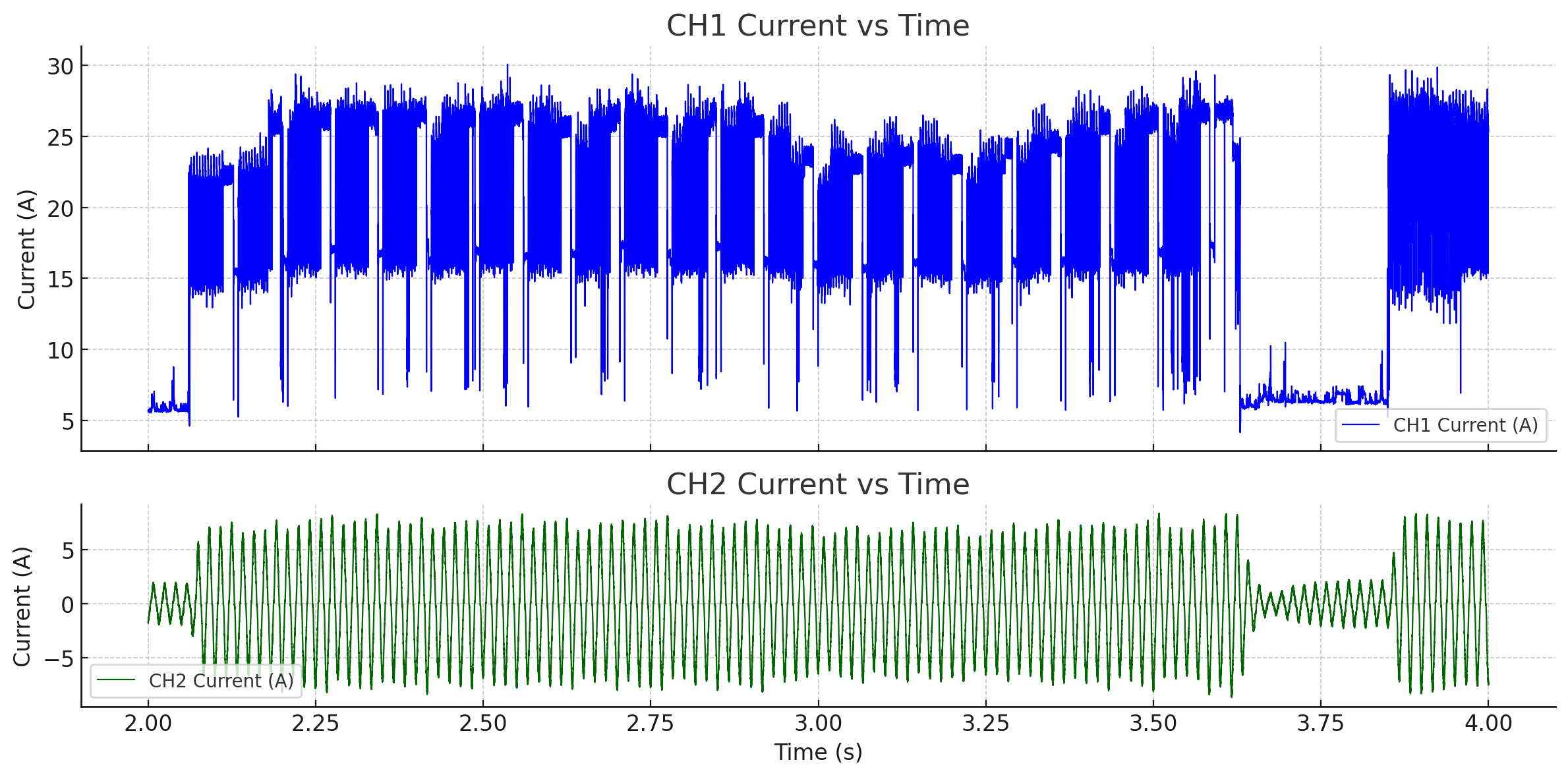}
    \caption{Longer time-scale capture showing repeated transitions.}
    \label{fig:psu_inertia_overall}
  \end{subfigure}
  \hfill
  \begin{subfigure}[b]{0.6\textwidth}
    \includegraphics[width=\textwidth]{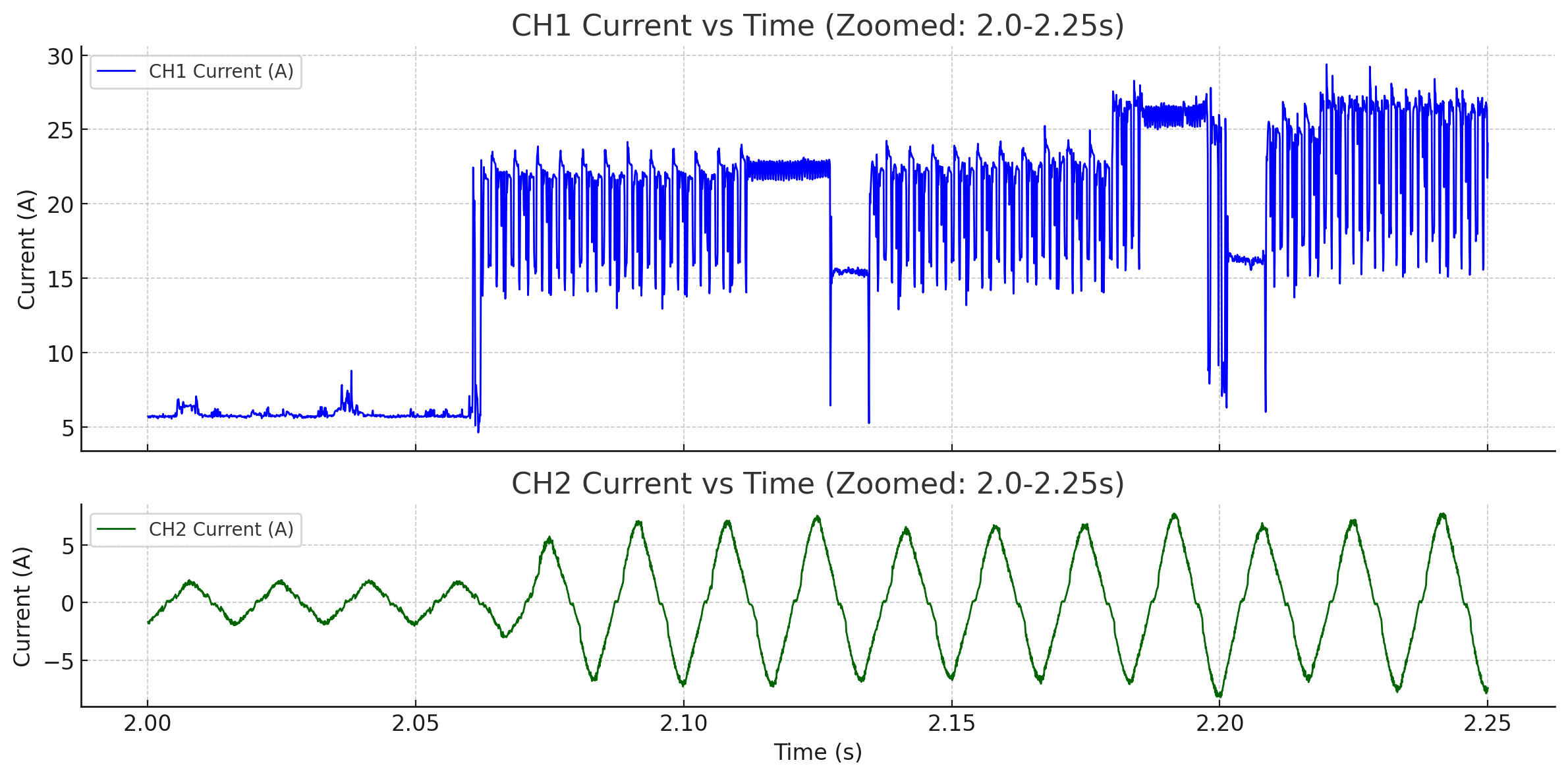}
    \caption{Zoomed view of the GPU current stepping from $\sim$5\,A to 20+\,A.}
    \label{fig:psu_inertia_zoom1}
  \end{subfigure}
  \hfill
  \begin{subfigure}[b]{0.6\textwidth}
    \includegraphics[width=\textwidth]{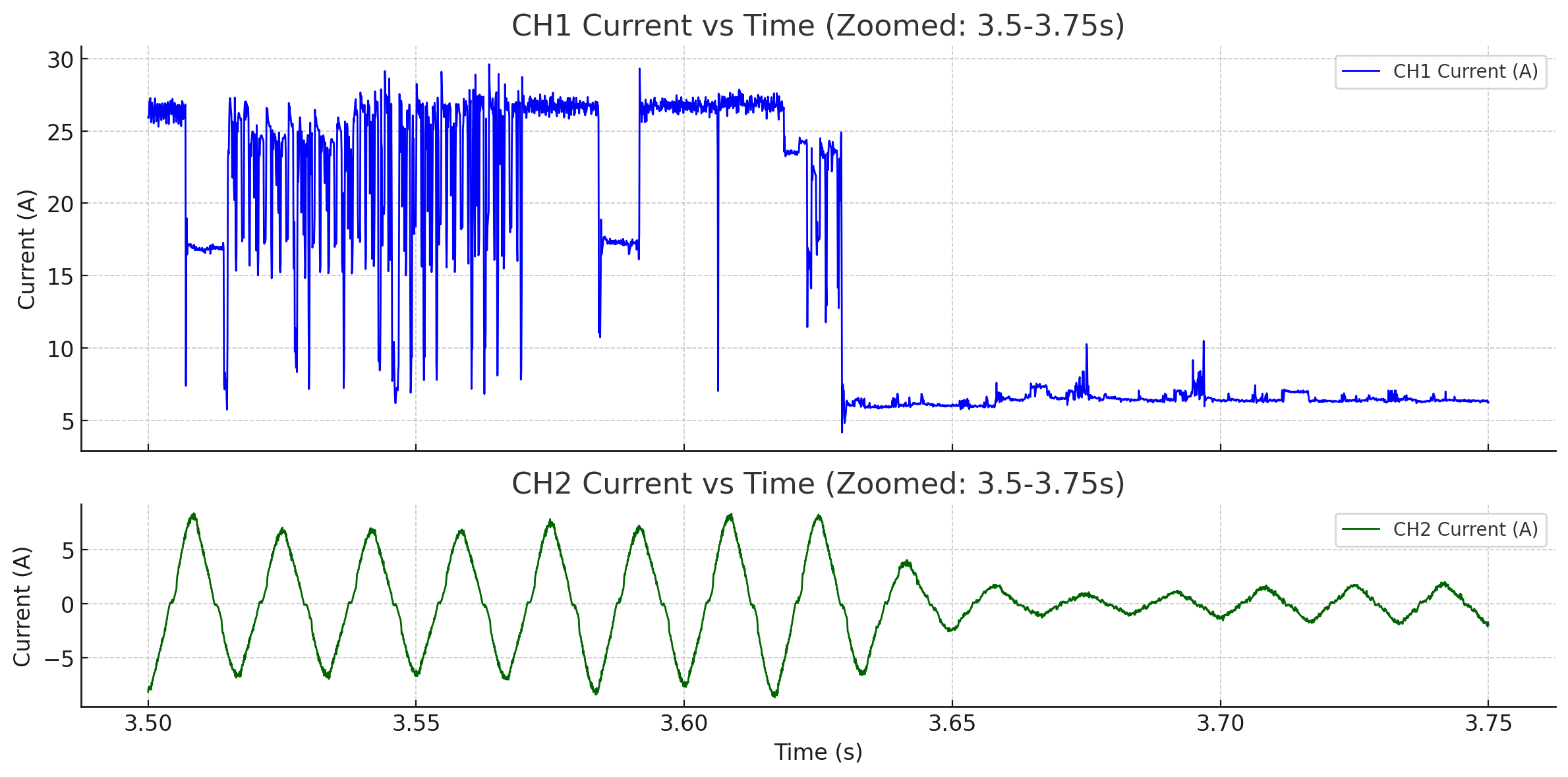}
    \caption{Further zoom on the negative transient region (3.5--3.75\,s).}
    \label{fig:psu_inertia_zoom2}
  \end{subfigure}
  \caption{Empirical PSU inertia example from single-phase 120\,V supply to a 12\,V GPU rail.
  (a)~A rapid load step on the GPU side, while the PSU input current lags by tens of milliseconds.
  (b)~An extended capture of multiple load surges.
  (c)~Closer look at negative transients, illustrating PSU hold-up capability.}
  \label{fig:psu_inertia_example}
\end{figure*}

While the preceding sections have presented a theoretical view of data center power dynamics, it is instructive to examine a concrete example of how a single power supply unit (PSU) buffers abrupt GPU load transitions in practice. Such ``inertia,'' on the order of a few tens of milliseconds, can be directly measured by comparing the GPU’s motherboard (DC) current against the PSU’s single-phase AC input current. Figures~\ref{fig:psu_inertia_zoom1}, \ref{fig:psu_inertia_overall}, and \ref{fig:psu_inertia_zoom2} showcase distinct operating points of the single-phase PSU feeding a GPU motherboard. The consolidated view in Figure~\ref{fig:psu_inertia_example} demonstrates the tens-of-milliseconds lag in the PSU’s AC input current whenever the GPU load steps rapidly on the 12V rail. The example provided here features a 1\,kW Gold-rated PSU powering a 
single NVIDIA-class GPU, but the same principles hold for larger AI compute nodes. 

\textbf{Single-PSU Measurement and Analysis}

\paragraph{Observed Load Step (GPU Rail).}
In Fig.~\ref{fig:psu_inertia_example}, the GPU rail current, $I_{\mathrm{GPU}}$, 
abruptly increases from around 5\,A to nearly 20\,A. Assuming a 12\,V rail, 
this corresponds to a load-step of roughly $180\,\mathrm{W}$. 
The GPU transitions between partial idle and active compute within a few milliseconds.

\paragraph{PSU ``Lag'' on the AC Input.}
The PSU’s AC input current, $I_{\mathrm{in}}$, does \emph{not} mirror 
this abrupt 180\,W rise in real time. Instead, it remains near its old amplitude 
for about 1--2 AC cycles (20--40\,ms) before ramping to match the new load. 
In some captures, this lag may extend to $\sim$50\,ms, indicating 
that the PSU’s internal bus capacitors and control loops are providing 
the extra energy (or absorbing it, in the case of a negative step). 
This timescale effectively defines the PSU’s ``inertia.''

\paragraph{Estimate of Stored Energy.}
We can approximate how many joules the PSU’s internal capacitors 
deliver during that 30--50\,ms mismatch by computing
\begin{equation}
  \Delta E \,\approx\, ( \Delta P ) \times ( \Delta t )
  \;=\; (180\,\mathrm{W}) \,\times\, (0.05\,\mathrm{s})
  \;=\; 9\,\mathrm{J}.
\end{equation}
This estimate aligns well with typical bulk-capacitor values in 
1--2\,kW server or workstation PSUs, which often store on the order of 5--15\,J. For sudden load \emph{drops}, this same buffer must 
absorb or shunt the excess energy to prevent a bus-voltage spike.

\textbf{Scaling to Rack-Level and Beyond}

Although this measurement involves a single GPU + single PSU, 
the fundamental notion of a $\sim$50\,ms buffering timescale generalizes 
to larger AI systems:

\begin{enumerate}[label=\roman*)]
    \item \textbf{Parallel Power Shelves:} 
    In a rack consuming 10--350\,kW, several PSUs or power shelves 
    (often 5--15\,kW each) operate in parallel. 
    Each still features tens of milliseconds of ``local inertia.'' 
    However, the \emph{aggregate} load step from multiple GPUs 
    can easily climb to tens or hundreds of kilowatts if many GPUs 
    initiate or end a compute phase simultaneously.

    \item \textbf{Aggregated Joule Demand:} 
    While one PSU might handle a 9\,J transient gracefully, 
    a synchronized load transition across 10 PSUs would require 90\,J, 
    and a larger row or entire data hall might demand thousands of joules 
    over the same short interval. If the sum of all PSU buffers 
    is insufficient, large upstream voltage swings or nuisance trips 
    can result.

    \item \textbf{Dominant Role of the Final Stage:} 
    As emphasized in Section~\ref{sec:power_chain_dynamics}, 
    these ``rack-level'' PSUs or power shelves are effectively 
    the final power-conversion stage, bridging the data center’s 
    208--480\,V supply (or MVAC in some topologies) and the GPUs. 
    Their 30--50\,ms bandwidth thus bounds the fastest net power 
    swing the upstream infrastructure must handle. Consequently, 
    multi-rack concurrency can still produce large sub-second transients 
    that propagate back to the utility interface.

\end{enumerate}

\textbf{Need for Additional Buffers and Controls}

Although tens-of-milliseconds smoothing is useful for \emph{individual} GPU load steps, it may not suffice 
when large clusters create near-simultaneous power surges in the multi-kW or multi-MW range. Therefore, 
operators typically deploy additional \textit{fast} energy storage or advanced scheduling, including:

\begin{itemize}[leftmargin=*]
    \item \textbf{High-Rate Energy Storage:} 
    Supercapacitors or specialized battery systems can absorb or deliver 
    multiple kilowatts (or megawatts) for a few seconds, bridging 
    any shortfall that the PSUs alone cannot handle.

    \item \textbf{Staggered Workload Coordination:} 
    HPC or AI schedulers may stagger checkpoints or batch boundaries so that 
    not all GPUs ramp simultaneously, distributing the load transitions 
    over a few hundred milliseconds to avoid huge instantaneous steps.

    \item \textbf{Enhanced UPS Solutions:} 
    UPS systems with high slew-rate designs and partial supercapacitor banks 
    can further buffer large negative or positive transients, 
    minimizing voltage excursions on the facility bus.
\end{itemize}

Without these additional measures, large data centers risk transient overvoltages, 
under-voltages, or excessive flicker at the PCC with the utility, 
ultimately threatening system reliability and potentially violating grid standards. 
Thus, while a single PSU’s 30--50\,ms inertia is sufficient to smooth out microsecond-level 
GPU transients, \emph{scaling up} to multi-rack or multi-megawatt loads demands supplementary 
power-engineering solutions to maintain stable, compliant operation. 


\subsection{Practical Considerations}

\subsubsection{UPS Battery Limitations}
Many UPS systems are designed for discharge over minutes, rather than for rapid charging over a span of milliseconds. Because their charge current is typically limited, a large negative transient can quickly lead to a DC-link overvoltage when the load drops abruptly. To mitigate such scenarios, several high-rate energy-absorption methods can be employed. One option is to incorporate supercapacitors, which are capable of accepting a sudden influx of energy thanks to their high charge acceptance. Another possibility is to deploy dedicated dump resistors or braking modules that can safely dissipate excess power when a load drop occurs. A further alternative is to use enhanced bi-directional batteries that can tolerate short bursts of high-current charging without compromising battery health. In practice, these strategies must be coordinated through a battery management system (BMS) that can handle rapid switching between discharge and charge modes; however, thermal management and cycle-life considerations often complicate the design and control of such BMS solutions.

\subsubsection{Coordination of Multiple Stages}
Large-scale data centers commonly rely on multiple AC/DC modules operating in parallel—often in the range of 50--100\,kW each—to reach total capacities of several megawatts. Although each module handles only a fraction of the total load, rapid load changes can still cause localized voltage transients if not carefully coordinated. Supervisory control layers typically manage these parallel modules by adjusting reference signals to maintain balanced bus voltages and avoid overstressing any single unit. In some cases, advanced scheduling or load shedding policies are employed to stagger large GPU load transitions, thereby reducing the magnitude of simultaneous ramp-up or ramp-down events. This approach helps maintain more stable power delivery and alleviates sudden demand surges that could exceed the module-level or facility-level bandwidth limits.

\subsubsection{Analytical Sizing and Stability}
Designing robust energy storage is a crucial aspect of preventing unwanted voltage excursions. For example, if a data center requires a hold-up time \( t_{\mathrm{hold}} \) at a peak power draw \( P_{\mathrm{peak}} \), the necessary bus capacitance \( C \) can be estimated by
\begin{equation}
    C \;\ge\; \frac{2\,P_{\mathrm{peak}}\,t_{\mathrm{hold}}}
                    {\,V_{\mathrm{dc}}^2 - (\,V_{\mathrm{dc}} - \Delta V\,)^2}.
\end{equation}
A larger capacitance can help manage voltage droop during positive load steps, but negative transients also become a factor, since the allowable voltage rise \(\Delta V_{\mathrm{up}}\) dictates how much energy storage or absorption capability is needed. Similar constraints apply to supercapacitors or batteries, where the maximum charge rate determines how rapidly they can accept surplus power during load drops.  

Control stability, on the other hand, hinges on maintaining adequate phase margin in each power stage, particularly at high load levels. Although raising the crossover frequency can improve transient response, pushing it too high risks destabilizing the overall system—especially when multiple modules are cascaded. Consequently, industry practice often sets conservative gain and phase margins (for example, \(\geq 60^\circ\)) at or near peak load to ensure that the power chain does not become oscillatory during abrupt GPU workload transitions. This trade-off between fast response and stable operation underscores the significance of “final-stage” bandwidth: no matter how rapidly downstream VRMs can respond, the slower dynamics of large upstream converters will ultimately bound the speed at which load variations can be accommodated safely.

\section{Discussions and Outlooks}
\label{sec:discussions_recommendations}

\subsection{Quantifying the Mismatch Between Load Dynamics and Energy Storage System (ESS) Capabilities}
\label{subsec:ess_mismatch}

It is important to note how different energy storage technologies, such as capacitors, supercapacitors, lithium-ion batteries, flow batteries, and diesel generators, each operate most effectively over particular discharge durations and power/energy levels. At one extreme, capacitors and supercapacitors deliver high power in sub-second intervals, making them suitable for very rapid load changes on the order of microseconds to milliseconds. At the other extreme, diesel generators (and similarly large engine-based systems) support moderate power for much longer durations, such as minutes to hours, thereby addressing extended outages or peak-shaving needs. AI accelerators like GPU-based clusters, however, often exhibit dynamic load patterns that include both rapid bursts of activity in the sub-millisecond range and multi-second surges for collective tasks. Consequently, a single energy storage solution generally cannot optimize performance across all these time and power scales.

To quantify whether a particular ESS technology can adequately meet a load transient, it is helpful to integrate the power difference over the relevant time period. Specifically, the metric
\begin{equation}
\Delta E_{\text{mismatch}}(t) 
= \int_{0}^{t} \bigl[ P_{\mathrm{demand}}(\tau) - P_{\mathrm{supply}}(\tau) \bigr] \, d\tau
\label{eq:delta_e_mismatch}
\end{equation}
captures how much extra energy is required (when \(\Delta E_{\text{mismatch}}(t) > 0\)) beyond what the ESS can deliver. Fast load transients, such as those common in GPU workloads, can lead to significant short-term mismatches if the storage element cannot supply (or absorb) power rapidly enough. In such scenarios, data-center designers may need to rely on a combination of storage technologies or fall back on alternative supplies to prevent large voltage dips or spikes. By analyzing \(\Delta E_{\text{mismatch}}(t)\) against the capabilities of capacitors, batteries, or other storage options, it becomes feasible to identify where a single ESS type falls short and where hybrid or tiered energy storage arrangements are necessary.

\subsection{Implications for Large-Scale GPU Deployments}

In large data centers that house thousands of GPUs, the load profile can feature a wide range of transient behaviors. Fast bursts on the order of microseconds to milliseconds often reach tens of kilowatts or more per server rack; to buffer these rapid events locally, operators rely on supercapacitors or low-inductance busbars that supply or absorb sudden current surges. Meanwhile, multi-second spikes in demand can occur during collective GPU tasks such as synchronized AI training steps or HPC job orchestration, where rack-level power consumption may exceed nominal levels by a substantial margin. If the grid or front-end PFC stage cannot respond quickly enough, batteries or large-scale flow batteries at the facility level are expected to cover this mismatch. For even longer disruptions, typically lasting minutes to hours, diesel or gas generators come into play, though these generators have comparatively slower response times and may not address sudden load transients on very short timescales. By carefully computing \(\Delta E_{\text{mismatch}}(t)\) based on typical workload and ramp profiles, data-center designers can pinpoint how much of the load must be supported by different energy-storage tiers. 

\subsection{Recommendations for Design and Integration}

Hybrid energy storage architectures are emerging as a robust solution to the time-scale mismatch problem. One effective strategy combines supercapacitors, which excel at providing or absorbing power surges that last only microseconds to a few seconds, with larger battery banks that can sustain multi-second bridging. On-site diesel or gas generators then serve extended outages or peak-shaving operations. Dynamically allocating the portion of load each storage layer handles requires a fast supervisory controller that engages supercapacitors for very short bursts and shifts responsibility to battery storage if the demand persists. At the same time, matching the slew rate of the final-stage converter to these storage devices is essential. This entails using topologies such as totem-pole or bidirectional PFC configurations and selecting semiconductor switches rated to handle large negative transients without damage.

Sizing each energy storage layer effectively involves applying the mismatch analysis to expected load scenarios. If short bursts exceed the cumulative storage or discharge capability of local bus capacitors, then designers must ensure supercapacitors or similarly high-rate solutions can capture that surge. Likewise, if multi-second intervals surpass battery capacity, an additional layer of storage or on-site generation is needed. Finally, the ability to expand infrastructure incrementally is critical as HPC and AI clusters grow in computational and power demands. By designing battery cabinets, supercapacitor modules, and distribution buses in a modular fashion, data centers can adapt to increasing load envelopes with minimal disruption to existing operations. In essence, carefully quantifying the dynamic mismatch between load and supply (via Eq.~\ref{eq:delta_e_mismatch}) and distributing storage responsibilities among multiple ESS tiers yields both improved resilience and greater efficiency, ensuring that sudden load ramps in GPU clusters do not compromise system performance or reliability.

\section{Conclusions}
\label{sec:conclusions}

This paper has examined the critical interplay between AI workload dynamics and power electronics in modern data centers, demonstrating that final-stage power conversion characteristics often constitute the fundamental bottleneck in system response capabilities. Through empirical measurements and theoretical analysis, we have shown that typical PSU implementations exhibit a built-in "inertia" that limits power system adaptability, regardless of downstream VRM performance. Our investigation reveals that GPU clusters create uniquely challenging load patterns—from microsecond-level inference transients to sustained training-related power swings—that push conventional power architectures beyond their original design parameters. While individual VRMs achieve microsecond responses, the cascade of conversion stages in data center power chains remains constrained by upstream converter bandwidth, typically in the kilohertz range. This limitation becomes particularly acute in large-scale deployments where synchronized GPU operations can create facility-wide power transients. Looking ahead, emerging solutions incorporating wide-bandgap devices, advanced control schemes, and hybrid energy storage systems show promise in bridging the growing gap between AI workload dynamics and power delivery capabilities—a critical consideration as data centers continue scaling to meet unprecedented computational demands.

\bibliographystyle{IEEEtran}
\bibliography{ref.bib}

\end{document}